\documentclass[pdflatex,sn-mathphys-num]{sn-jnl}

\usepackage{graphicx}%
\usepackage{multirow}%
\usepackage{amsmath,amssymb,amsfonts}%
\usepackage{amsthm}%
\usepackage{mathrsfs}%
\usepackage[title]{appendix}%
\usepackage{xcolor}%
\usepackage{textcomp}%
\usepackage{manyfoot}%
\usepackage{booktabs}%
\usepackage{algorithm}%
\usepackage{algorithmicx}%
\usepackage{algpseudocode}%
\usepackage{listings}%

\theoremstyle{thmstyleone}%
\newtheorem{theorem}{Theorem}
\theoremstyle{thmstyletwo}%

\theoremstyle{thmstylethree}%
\newtheorem{definition}{Definition}%

\usepackage[numbers]{natbib}
\usepackage{tikz}          
\usetikzlibrary{quantikz}  
\usepackage{bbm}
\usepackage{bbold}
\usepackage{relsize}
\usepackage{comment}

\usepackage{booktabs}

\newcommand{\field}[1]{\mathbb{#1}}
\newcommand{\N}{\field{N}}

\newcommand{\R}{\field{R}}
\newcommand{\C}{\field{C}}
\newcommand{\hilbert}[1]{\mathcal{H}_{#1}}
\newcommand{\tr}{\operatorname{tr}}

\begin{document}

\title[Semidefinite Programming for Optimal Quantum Cloning: A Computational Framework]{Semidefinite Programming for Optimal Quantum Cloning: A Computational Framework}


\author[]{\fnm{Jörg} \sur{Hettel}}\email{joerg.hettel@hs-kl.de}

\affil[]{\orgdiv{Department of Computer Science and Microsystems Technology}, \orgname{University of Applied Sciences Kaiserslautern}, \orgaddress{\country{Germany}}}


\abstract{While algebraic derivations establish theoretical limits for quantum cloning, practical implementations require explicit operator representations that are often unavailable analytically. We present a computational framework that reformulates cloning optimization as a search over completely positive trace-preserving maps using the Choi-Jamiołkowski isomorphism and Semidefinite Programming.
The framework (i) numerically certifies global optimality through primal-dual strong duality and (ii) automatically extracts operational Kraus operators from the optimal Choi matrix via spectral decomposition. We systematically treat universal, phase-covariant, asymmetric, and entanglement cloning scenarios, providing—for the first time—a unified  computational catalogue of explicit, implementable Kraus representations across all major cloning families, including higher-order processes and arbitrary input state distributions.
As an application, we analyse optimal cloning attacks on BB84 under depolarizing noise, demonstrating how the extracted operators enable quantitative security analysis in realistic noisy quantum channels. An open-source implementation enables community validation and extension.}


\keywords{Quantum Cloning, Semidefinite Programming, Choi-Jamiołkowski Isomorphism, Kraus Operators. }

\maketitle

\section{Introduction}

The no-cloning theorem \cite{WoottersZurek1982} establishes a fundamental boundary in quantum information: an unknown quantum state cannot be perfectly duplicated. Despite this fundamental limitation, approximate cloning plays a crucial role in quantum cryptography and information processing. Optimal cloning strategies are essential for understanding eavesdropping attacks on quantum key distribution (QKD) \cite{Gisin2002} and for assessing the limits of quantum state broadcasting.

The earliest families of such cloners were the universal $1 \mapsto2$ qubit cloners, formulated by Bužek and Hillery \cite{Buzek1996}, and later mathematically characterized by Werner \cite{Werner1998}, who connected their structure to irreducible SU(2) representations and the symmetrization of tensor powers.

The extension to universal $M \mapsto N$ cloners was developed by Keyl and Werner \cite{Keyl1999}, whose representation-theoretic formulation based on Schur–Weyl duality yields closed expressions for the optimal cloning fidelity. Since then, various restricted or symmetry-reduced cloning transformations have been explored, including phase-covariant cloning of equatorial qubit states \cite{Bruss2000,Ariano2003}, asymmetric cloning \cite{Cerf2000a,Cerf2000b,Cerf2000c} and entanglement cloning \cite{Lamoureux2004}. A detailed general overview of the various variants can be found in Fan et al. \cite{Fan2014}. These analytically solvable models provide valuable insight into the structure and limitations of physically admissible cloning channels.

Parallel to these conceptual developments, semidefinite programming (SDP) has emerged as one of the most powerful techniques for quantum information optimization problems. The Choi–Jamiołkowski isomorphism enables the formulation of channel optimization tasks as SDPs with linear matrix inequalities, facilitating provable optimality certificates through primal–dual bounds. SDPs now play a central role in entanglement detection \cite{Navescues2009}, quantum state discrimination \cite{Chefles2000}, channel certification \cite{Piani2009}, and specific applications such as quantum money forgery \cite{Molina2013}. Early contributions by Audenaert \& De Moor \cite{Audenaert2002}, and Watrous \cite{Watrous2018} demonstrated that SDP techniques can efficiently handle high-dimensional quantum problems that lack closed-form analytic solutions.

Our work combines these two strands of research.
We provide a complete computational pipeline for various quantum cloning processes, serving as a computational tutorial. We integrate the Choi-Jamiołkowski isomorphism into a systematic SDP framework that not only calculates optimal fidelities but also extracts operational Kraus operators through spectral decomposition. Thus, our contribution is threefold: (i) the automated extraction of operational Kraus representations for cloning operations, (ii) the systematic treatment of arbitrary state distributions beyond symmetric cases, and (iii) a ready-to-use computational pipeline demonstrated through cryptographic applications.

The remainder of this article is organized as follows: Section \ref{sec:preliminaries} establishes the necessary mathematical foundations, focusing on the Choi-Jamiołkowski isomorphism. Section \ref{sec:semidefinite_programming} details the semidefinite programming formulation, including the primal-dual structure and certification. The extraction and validation of Kraus operators from the obtained Choi matrix are presented in Section \ref{sec:kraus_extraction}, and Section \ref{sec:results} provides a comprehensive evaluation of our framework across various cloning scenarios. A cryptographic application—a cloning attack on the BB84 protocol accounting for noise—is discussed in Section \ref{sec:BB84Attack}. Finally, Section \ref{sec:discussion} discusses the results and outlines potential extensions, and Section \ref{sec:conclusion} summarizes our findings.

\section{Mathematical Preliminaries} \label{sec:preliminaries}
We provide a self-contained review of quantum channel theory essential for the SDP formulation. Standard references for this formalism include \cite{Wilde_2017, Watrous2018}.

A quantum channel is a completely positive, trace-preserving (CPTP) linear map $\mathcal{E}(\rho_{\text{in}})$ acting on density matrices ($\rho_{in} \in \mathcal{L}(\hilbert{in})$):

\begin{definition}[Quantum Channel]
    A quantum channel is a linear map $\mathcal{E}: \mathcal{L}(\hilbert{\text{in}}) \to \mathcal{L}(\hilbert{\text{out}})$ between finite-dimensional complex Hilbert spaces that is:
    \begin{enumerate}
        \item \textit{Trace-Preserving:} $\tr[\mathcal{E}(\rho)] = \tr[\rho]$ for all density operators $\rho$.
        \item \textit{Completely Positive:} $(\mathbb{1}_n \otimes \mathcal{E})(\sigma) \succeq 0$ for all $\sigma \succeq 0$ and $n \in \N$, where $\mathbb{1}_n$ is the identity on $\C^n$.
    \end{enumerate}
\end{definition}
Throughout this work, we denote the input and output dimensions as $d_{\text{in}}$ and $d_{\text{out}}$, respectively. For a state $\ket{\phi} = \sum_k c_k \ket{k}$, its complex conjugate is defined as $\ket{\phi^*} = \sum_k c_k^* \ket{k}$ in the computational basis.

Every CPTP map admits a Kraus representation:
\begin{equation}\label{eq:kraus}
    \mathcal{E}(\rho) = \sum_{k=1}^{r} K_k \rho K_k^\dagger,
\end{equation}
where $\{K_k\}_{k=1}^r$ satisfy the completeness relation $\sum_k K_k^\dagger K_k = \mathbb{1}_{\text{in}}$ and $k$ is the Kraus rank. For a $1 \to 2$ qubit cloner, $K_k \in \C^{4 \times 2}$ maps the input space $\hilbert{\text{in}} = \C^2$ to the output space $\hilbert{\text{out}} = \C^2 \otimes \C^2$.

The Choi representation of a quantum channel plays a central role in our discussion:
\begin{definition}[Choi Matrix]
    The Choi matrix of a linear map $\mathcal{E}: \mathcal{L}(\hilbert{\text{in}}) \to \mathcal{L}(\hilbert{\text{out}})$ is
    \begin{equation}\label{eq:choi_def}
        J(\mathcal{E}) = (\mathbb{1}_{\text{in}} \otimes \mathcal{E})\left( |\Gamma \rangle \langle \Gamma | \right),
    \end{equation}
    where $ |\Gamma\rangle = \sum_{j=0}^{d_{\text{in}}-1} |j\rangle \otimes |j\rangle$ is the unnormalized maximally entangled state.
\end{definition}
Note that $J(\mathcal{E}) \in \mathcal{L}(\hilbert{\text{in}} \otimes \hilbert{\text{out}})$ is a positive semi-definite operator with $\tr J(\mathcal{E}) = d_{\text{in}}$  (due to the unnormalized state).

An alternative and computationally convenient way to construct the Choi matrix is through its action on the basis elements of the input space. Given an orthonormal basis $\{ \ket{i}\}$ for the $d_{\text{in}}$-dimensional input space and $\{ \ket{j}\}$ for the $d_\text{out}$-dimensional output space, the Choi matrix can be expressed as
\begin{equation}
    J(\mathcal{E}) = \sum_{i,j} \ket{i}\bra{j} \otimes \mathcal{E}\left( \ket{i}\bra{j} \right).
\end{equation}
This formulation reveals the block structure of the Choi matrix explicitly: the $(i,j)$-th block is given by $\mathcal{E} ( \ket{i}\bra{j} )$, which is a $d_\text{out} \times d_\text{out}$ matrix. Thus,  $\ket{i}\bra{j}$ is a $d_\text{in} \times d_\text{in}$ block matrix where each block is a $d_\text{out} \times d_\text{out}$ matrix, yielding an overall dimension of $(d_\text{in} \cdot d_\text{out}) \times (d_\text{in} \cdot d_\text{out})$. This construction is particularly useful for numerical implementation.

\subsection{Choi-Jamiołkowski Isomorphism}
The Choi-Jamiołkowski isomorphism establishes a bijection between linear maps $\mathcal{E}(\rho)$ and the Choi operators $J(\mathcal{E})$. CPTP maps have the following properties \cite{Watrous2018}:

\begin{theorem}[CPTP Characterization]\label{thm:cptp_choi}
    A linear map $\mathcal{E}$ is CPTP if and only if its Choi matrix $J(\mathcal{E})$ satisfies:
    \begin{enumerate}
        \item $J(\mathcal{E}) \succeq 0$ (complete positive)
        \item $\tr_{\text{out}}[J(\mathcal{E})] = \mathbb{1}_{\text{in}}$ (trace preservation)
    \end{enumerate}
\end{theorem}

\subsection{Fidelity Functional}
To evaluate the performance of a quantum cloning channel, we employ the fidelity measure. For a (possibly mixed) quantum state $\rho$ and an ideal pure target state $\ket{\Psi}$, the fidelity is defined as:
\begin{equation}
    F = \bra{\Psi} \rho \ket{\Psi}. 
\end{equation}
It quantifies the overlap between the two states, ranging from $0$ for orthogonal states to $1$ for identical states. When $\rho$ is the output of a quantum channel $\mathcal{E}$ acting on an input state $\rho_{in} = \ket{\phi}\bra{\phi}$, the fidelity becomes:
\begin{equation}
    F = \bra{\Psi} {\cal E}(\ket{\phi}\bra{\phi}) \ket{\Psi}. 
\end{equation}
Using the properties of the Choi-Jamiołkowski isomorphism (see Appendix \ref{app:choi_derivation}), this expression can be reformulated in terms of the Choi matrix $J(\mathcal{E})$:
\begin{equation}
    F =  \bra{\phi^*} \bra{\Psi}   J({\cal E}) \ket{\phi^*} \ket{\Psi} .
\end{equation}
By defining the operator
\begin{equation}
    \Omega =  \ket{\phi^*}\bra{\phi^*} \otimes \ket{\Psi}\bra{\Psi}, 
\end{equation}
we obtain the fidelity as a linear functional of the Choi matrix:
\begin{equation}
    F =  \tr \left[ J({\cal E})\cdot \Omega \right].
\end{equation}
Here, $\ket{\phi}$ denotes the input state of the channel, while $\ket{\Psi}$ represents the desired target state (e.g., the ideal clones).

\section{Semidefinite Programming Formulation} \label{sec:semidefinite_programming}
Semidefinite programming (SDP) is a class of convex optimization problems that is mathematically well-established \cite{Boyd2004}. SDP has numerous applications within quantum computing and quantum information theory (see, e.g., \cite{Audenaert2002, Watrous2018, Siddhu2022, Skrzypczyk2023, Mironowicz2024}).
We formulate optimal quantum cloning as a convex optimization problem over the space of CPTP maps to maximize the average fidelity. To this end, we consider a set of input states $\mathcal{S} = \{ \ket{\psi_k} \}$ and their corresponding ideal output states. For an $M \to N$ cloning channel, the average performance is captured by the target operator $\Omega$:
\begin{equation}
    \Omega = \frac{1}{|\mathcal{S}|} \sum_{|\psi_k\rangle \in \mathcal{S}} \left( \ket{\psi^*_k} \bra{\psi^*_k}\right)_\text{in}^{\otimes M} \otimes \left( \ket{\psi_k} \bra{\psi_k} \right)_{\text{out}}^{\otimes N}.  
\end{equation}

\begin{figure}[h]
    \centering
    \includegraphics[height=3cm]{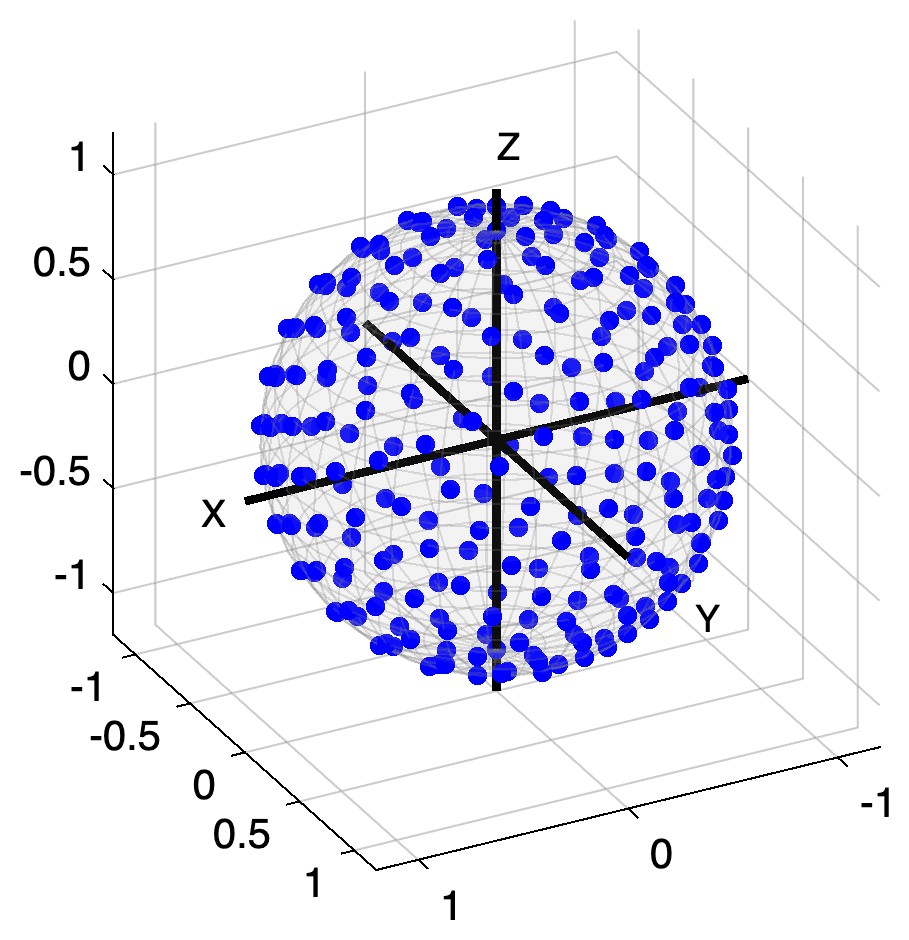}\hspace{3 mm}
    \includegraphics[height=3cm]{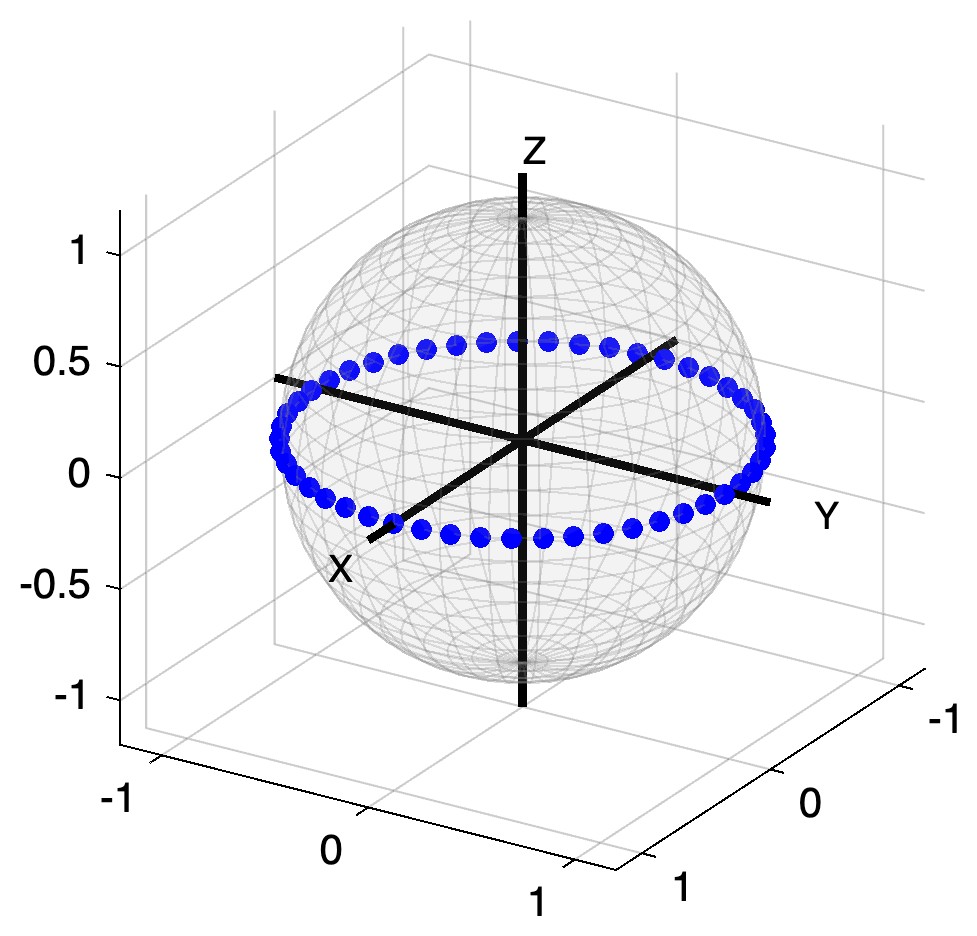}\hspace{3 mm}
    \includegraphics[height=3cm]{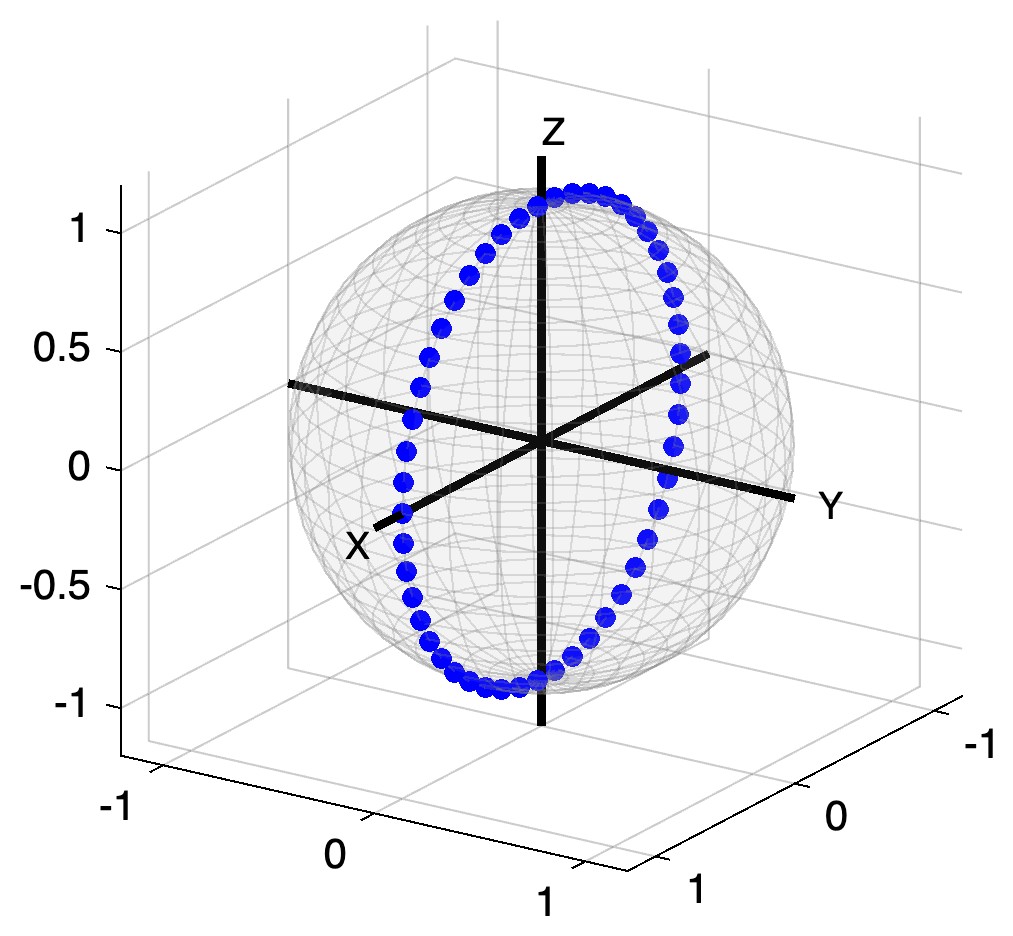}
    \caption{Various qubit distributions on the Bloch sphere used for numerical optimization.}\label{QubitDistribution}
\end{figure}

Figure \ref{QubitDistribution} illustrates different distributions of states $\{\ket{\psi_k}\}$ on the Bloch sphere that constitute the sampling set $\mathcal{S}$ for $\Omega$.

In the following, we discuss the SDP formulation for symmetric $1 \to 2$ cloning in detail. The extension to the general $M \to N$ case can be performed in a systematic manner.

\subsection{The Sampling Set}\label{sec:sampling_set}
For a symmetric $1 \to 2$ cloning channel, the global target operator is defined as:
\begin{equation}
    \Omega_{\text{global}} = \frac{1}{|\mathcal{S}|} \sum_{|\psi_k\rangle \in \mathcal{S}} \left( \ket{\psi^*_k} \bra{\psi^*_k}\right)_{\text{in}} \otimes \left( \ket{\psi_k} \bra{\psi_k} \right)_{\text{out}_A} \otimes \left( \ket{\psi_k} \bra{\psi_k} \right)_{\text{out}_B}.
\end{equation}
By choosing an appropriate sampling set $\mathcal{S}$ and maximizing the functional $\tr[J \Omega_{\text{global}}]$, we obtain the optimal global fidelity:
\begin{equation}
    F_{\text{global}} = \bra{ \phi_{in} \otimes \phi_{\text{in}}} \tr \left[ J \cdot \Omega_{\text{global}} \right]  \ket{ \phi_{\text{in}} \otimes \phi_{\text{in}}} = \tr \left[ J \cdot \Omega_{\text{global}} \right]
\end{equation}
To determine the optimal single-copy fidelity (local fidelity), we consider the performance relative to a single output, for instance, subsystem $A$. In this case, we define a local target operator:
\begin{equation}
    \Omega_{A} = \frac{1}{|\mathcal{S}|} \sum_{|\psi_k\rangle \in \mathcal{S}} \left( \ket{\psi^*_k} \bra{\psi^*_k}\right)_{\text{in}} \otimes \left( \ket{\psi_k} \bra{\psi_k} \right)_{\text{out}_A}.
\end{equation}
By setting $\Omega_{\text{local}} = \Omega_A \otimes \mathbb{1}_{\text{out}_B}$, the optimization objective simplifies to:
\begin{equation}
    F_{\text{local}} = \tr [J (\Omega_A \otimes \mathbb{1}_{\text{out}_B})] = \tr [\Omega_A \text{Tr}_B(J)],
\end{equation}
effectively tracing out subsystem $B$ from the optimization objective.
To ensure that both clones have identical quality in a symmetric cloning scenario, we must impose an additional symmetry constraint on the SDP. This can be formulated as a permutation symmetry requirement on the Choi matrix:
\begin{equation}
    J = (\mathbb{1}_{\text{in}} \otimes S_{AB}) J (\mathbb{1}_{\text{in}} \otimes S_{AB})^\dagger,
\end{equation}
where $S_{AB}$ is the swap operator acting on the Hilbert spaces $\mathcal{H}_{\text{out}_A} \otimes \mathcal{H}_{\text{out}_B}$.

The optimization effectively performs a restricted form of channel tomography, making the choice of the sampling set essential. Whenever symmetric informationally complete (SIC) sets exist for the relevant Hilbert-space dimension, we use the corresponding normalized SIC states. SIC states provide uniform coverage of the state space, are informationally complete by construction, and minimize sampling redundancy.

If no SIC set is known or available in a given dimension, we instead employ uniformly distributed samples drawn from the Haar measure on pure states. This ensures unbiased coverage and asymptotic informational completeness. As a computationally efficient alternative, approximate complex projective designs may also be used, offering near-uniform sampling with significantly fewer states. These strategies guarantee that the SDP remains well constrained even in dimensions lacking an exact SIC structure. Specific examples can be found in appendix \ref{app:NumericalResults}.

In what follows, we focus on local fidelity for single copies and omit the index 'local’ on $\Omega$.

\subsection{The Primal Problem}
The primal formulation for the SDP provides a lower bound for the fidelity. The optimization seeks the Choi operator $J$ that maximizes average fidelity:
\begin{equation}\label{eq:primal_sdp}
    \begin{aligned}
        \text{maximize} \quad & \text{Tr}[J \Omega] \\
        \text{subject to} \quad & J \in \mathcal{L}(\hilbert{\text{in}} \otimes \hilbert{\text{out}}) \\
        & J \succeq 0, \\
        & \tr_\text{out}[J] = \mathbb{1}_\text{in}, \\
        & J = (\mathbb{1}_\text{in} \otimes S_{AB}) J (\mathbb{1}_\text{in} \otimes S_{AB})^\dagger \quad \text{(symmetry of the output)}.
    \end{aligned}
\end{equation}
The first constraint ensures the complete positivity of the map, while the second (partial trace) constraint enforces the trace-preserving property. The final constraint ensure permutation symmetry of the outputs when considering symmetric cloning. $S_{AB}$ denotes the swap operator $\mathbb{1}_\text{in} \otimes S_{AB}$ as defined previously.

\subsection{The Dual Problem}
The dual formulation provides an upper bound on the optimal fidelity. By introducing the Lagrange multipliers $Y$ (a Hermitian operator on $\mathcal{H}_{\text{in}}$ associated with the trace-preservation constraint) and $Z$ (a Hermitian operator on $\mathcal{H}_{\text{in}} \otimes \mathcal{H}_{\text{out}}$ associated with the symmetry constraint), the dual problem is given by:
\begin{equation}\label{eq:dual_sdp}
    \begin{aligned}
        \text{minimize} \quad & \tr[Y] \\
        \text{subject to} \quad & Y \in \mathcal{L}(\hilbert{\text{in}}), \quad Y = Y^\dagger \\
        & Z \in \mathcal{L}(\hilbert{\text{in}} \otimes \hilbert{\text{out}}), \quad Z = Z^\dagger \\
        & Y \otimes \mathbb{1}_\text{out} + (Z - S_{AB} Z S_{AB}^\dagger) \succeq \Omega \quad \text{(symmetry of the output)}.
    \end{aligned}
\end{equation}

\subsection{Numerical Certification and Strong Duality}
The duality gap, defined as the difference between the dual and primal objective values, can be used to certify the optimality of the numerical solution. For our formulation, the gap is given by:
\begin{equation}
    \Delta = \left| \text{Tr}(Y) - \text{Tr}(J \Omega) \right|.
\end{equation}
A gap approaching machine precision (near $\epsilon_{\text{mach}}$) confirms that the calculated cloning channel is indeed optimal. This indicates that strong duality holds, which is expected as the problem is a strictly feasible convex optimization satisfying Slater's condition:
\begin{equation}
    \max_J \text{Tr}[J \Omega] = \min_Y  \text{Tr}[Y].
\end{equation}

\section{Kraus Operator Extraction} \label{sec:kraus_extraction}

Once the optimal Choi matrix $J$ is obtained via the SDP, the physical action of the cloning channel can be represented in terms of its Kraus operators $\{K_i\}$. This representation is essential for characterizing the operational noise processes and for proposing concrete experimental implementations, such as quantum circuit decompositions.

\subsection{Spectral Decomposition Method}
The most direct route to obtaining the Kraus operators is through the spectral decomposition of the Choi matrix $J$. This procedure is standard in quantum channel theory \cite{Wilde_2017,Watrous2018}. Since $J$ is positive semidefinite by construction, it can be decomposed as:
\begin{equation}
    J = \sum_{i=1}^r \lambda_i |v_i\rangle \langle v_i|,
\end{equation}
where $\lambda_i > 0$ are the non-zero eigenvalues and $|v_i\rangle$ are the corresponding orthonormal eigenvectors. The Kraus operators $K_i$ can be reconstructed by reshaping these eigenvectors into matrices. Given our convention for the Hilbert space $\mathcal{H}_\text{in} \otimes \mathcal{H}_\text{out}$, the extraction follows:
\begin{equation}
    K_i = \sqrt{d_\text{in} \lambda_i} , \text{vec}^{-1}(|v_i\rangle),
\end{equation}
where $\text{vec}^{-1}$ denotes the inverse vectorization (reshaping) operation that maps a vector of dimension $d_{in} \cdot d_{out}$ to a $d_\text{out} \times d_\text{in}$ matrix. The number of non-zero eigenvalues $r$ corresponds to the Kraus rank of the channel, representing the minimum number of operators required to describe the transformation.

\subsection{Validation of Extracted Operators}
To ensure the numerical integrity of the extraction process, the resulting set of Kraus operators $\{K_i\}$ must be validated against the defining properties of a quantum channel. Specifically, we verify the completeness relation, which corresponds to the trace-preserving condition:
\begin{equation}
    \sum_{i=1}^r K_i^\dagger K_i = \mathbb{1}_{\text{in}}.
\end{equation}
Furthermore, we ensure that the action of the channel, $\mathcal{E}(\rho) = \sum_i K_i \rho K_i^\dagger$, yields the same fidelity results as those predicted by the SDP objective $\tr [J \Omega]$. This serves as a critical consistency check between the operator-sum representation and the Choi matrix optimization, confirming the robustness of the numerical pipeline.

\subsection{Unitary Freedom and Canonical Forms}
The Kraus representation is not unique; for any unitary matrix $U$, the set of operators $K'_i = \sum_j U_{ij} K_j$ represents the same quantum channel $\mathcal{E}$. In this work, we utilize the spectral decomposition of the Choi matrix to obtain the canonical Kraus representation. In this representation, the operators are orthogonal with respect to the Hilbert-Schmidt inner product:
\begin{equation}
    \tr (K_i^\dagger K_j) = \delta_{ij} \cdot (\text{const.}),
\end{equation}
where the specific normalization depends on the Choi matrix scaling. This canonical form can be used for identifying the dominant transformation modes of the cloner and simplifies the transition to a dilated unitary evolution via Stinespring's dilation theorem \cite{Watrous2018}.

\section{Numerical Results and Validation} \label{sec:results}
In this section, we present the numerical results obtained from our SDP-based optimization framework. We compare our findings with established analytical bounds to validate the precision of the implementation and explore various cloning scenarios.

\subsection{Computational Setup}
The optimization problems were implemented in MATLAB (Release R2025b Update 5) using the \textit{CVX} modeling framework (Version 2.2, Build 9) \cite{cvx}. The Choi matrices and cloning templates were constructed using the \textit{QETLAB} library \cite{qetlab}, which provides efficient routines for partial traces and permutation operators. We employed the \textit{SDPT3} solver (version 4) for all computations. Although SDPT3 is a primal-dual solver that solves both problems simultaneously and provides optimal values for both variables, the dual problem was also formulated and solved explicitly to verify the certification process. Due to computational resource constraints (primarily memory), the framework currently supports numerical simulations of systems with up to 8 qubits, distributed across the input and output registers. 

Execution times were determined using the MATLAB function \texttt{timeit} for automated benchmarking. All calculations were carried out on a standard workstation (MacBook Pro with M2 processor and 64 GB RAM).
All implementations are available on \href{https://github.com/hettel/QuantumCopyChannels}{GitHub} under the MIT license \cite{Hettel2026_Code}.

\subsection{Universal Cloning}
As a benchmark, we first consider the Universal Quantum Cloning Machine (UQCM), which is required to perform equally well for all states on the Bloch sphere. We compare our numerical findings with the results from \cite{Buzek1996, Keyl1999}.  By solving the primal and dual SDPs with the symmetry constraint, our results confirm the analytical fidelity bound:
\begin{equation}
    F_{\text{univ}}(1 \to N) = \frac{2N + 1}{3N}.
\end{equation}
To determine the optimal fidelity numerically, it is sufficient for $\Omega$ to be constructed from only the following four states:
\begin{equation} \label{eqn:sic-povm}
    \ket{\psi_0} = \ket{0}, \quad \ket{\psi_{1,2,3}} = \sqrt{\frac{1}{3}}\ket{0} + \sqrt{\frac{2}{3}} e^{i \frac{2\pi (k-1)}{3}}\ket{1}, \quad k=1,2,3.
\end{equation}
These states correspond to a Symmetric Informationally Complete POVM (SIC-POVM) and are informationally complete in the sense that they allow for the full reconstruction of any input state from measurement data \cite{Renes2004}.

If states are instead distributed uniformly (but randomly) across the Bloch sphere to construct $\Omega$, a larger number of samples is required to reach high numerical precision. For this case, the input states are sampled uniformly from the entire Bloch sphere (see Fig. \ref{QubitDistribution}, left). In fact, the approximation quality increases slowly with $\mathcal{O}(1/|\mathcal{S}|^2)$ (see Table \ref{tab:symmetric1to2cloning} in the appendix). 

The scalability of the SDP approach is demonstrated for different input and output Hilbert space dimensions. For producing identical copies, the symmetry constraint must be expanded to require that all marginal channels are equivalent. By making appropriate modifications to $\Omega$ and the symmetry conditions of the SDP, more general cloning processes can be analyzed. For a universal symmetric $M \to N$ cloner, the optimal fidelity is given by:
\begin{equation} \label{eqn:universal_cloning}
    F^{\text{opt}}_{M \to N} = \frac{M(N+1) + N}{N(M+2)}.
\end{equation}
Where two equally input qubits  (M = 2), an approximate SIC state space was used; for three input qubits (M = 3), sampling was performed uniformly across the input space. Table \ref{tab:MtoNcloner} in the appendix compares the theoretical and numerical results, showing excellent agreement with Eq.~\!(\ref{eqn:universal_cloning}).

From the optimized Choi matrix $J_{\text{opt}}$, the Kraus representation of the cloning channel can be extracted. In such cases, it is often possible to derive an algebraic expression from the numerical values, which can then be checked again. For the universal $1 \to 2$ cloner, we obtain the following Kraus operators:
\begin{equation}
    K_1 =  \frac{1}{\sqrt{6}} \begin{pmatrix}
        2 & 0 \cr
        0 & 1\cr
        0 & 1\cr
        0 & 0
    \end{pmatrix}, \qquad
    K_2 = \frac{1}{\sqrt{6}} \begin{pmatrix}
        0 & 0 \cr
        1 & 0 \cr
        1 & 0 \cr
        0 & 2
    \end{pmatrix}    
\end{equation}
For the $1 \to 3$ cloner, the following algebraic Kraus representation is easily derived from the numerical output:
\begin{equation}
    K_1 = \frac{1}{6} \left(\begin{array}{cc}
        3\,\sqrt{2} & 0\\
        0 & \sqrt{2}\\
        0 & \sqrt{2}\\
        0 & 0\\
        0 & \sqrt{2}\\
        0 & 0\\
        0 & 0\\
        0 & 0
    \end{array}\right), \qquad
    K_2 = \frac{1}{6} \left(\begin{array}{cc}
        0 & 0\\
        0 & 0\\
        0 & 0\\
        \sqrt{2} & 0\\
        0 & 0\\
        \sqrt{2} & 0\\
        \sqrt{2} & 0\\
        0 & 3\,\sqrt{2}
    \end{array}\right), \qquad
    K_3 = \frac{1}{3} \left(\begin{array}{cc}
        0 & 0\\
        1 & 0\\
        1 & 0\\
        0 & 1\\
        1 & 0\\
        0 & 1\\
        0 & 1\\
        0 & 0
    \end{array}\right)
\end{equation}
These numerical results align perfectly with the structured sparsity expected from the symmetry of the universal cloning transformation.

\subsection{Covariant Cloning}\label{sec:covariant}

To analyze covariant cloning \cite{Bruss2000}, the input states are restricted to a great circle of the Bloch sphere (see Fig. \ref{QubitDistribution}, center and right). While we obtain the same optimal fidelity for any great circle, the corresponding Kraus operators differ.

We consider $1 \mapsto N$ phase-covariant cloning of qubits of the form $\ket{\psi} = (\ket{0} + e^{i\phi}\ket{1})/\sqrt{2}$. To construct $\Omega$, it is sufficient to use a minimal informationally complete set consisting of three states forming an equilateral triangle on the equatorial plane:
\begin{equation}
    \ket{\psi_k} = \frac{1}{\sqrt{2}} \left( \ket{0} + e^{i \frac{2\pi(k-1)}{3}} \ket{1}\right), \quad k=1,2,3.
\end{equation}

For the $1 \to 2$ phase-covariant cloner, the SDP yields the following Kraus operators:
\begin{equation} \label{eqn:phase_covariant}
    K_1 =  \frac{1}{2} \begin{pmatrix}
        \sqrt{2} & 0 \cr
        0 & 1\cr
        0 & 1\cr
        0 & 0
    \end{pmatrix}, \qquad
    K_2 = \frac{1}{2} \begin{pmatrix}
        0 & 0 \cr
        1 & 0 \cr
        1 & 0 \cr
        0 & \sqrt{2}
    \end{pmatrix}. 
\end{equation}
Next, we consider qubits restricted to the $x$-$z$ plane (real amplitudes) of the form $\ket{\psi} = \cos\phi \ket{0} + \sin\phi \ket{1}$. Using a similar sampling set for $\Omega$, the SDP extracts:
\begin{equation}
    V_1 =  \begin{pmatrix}
        \frac{1}{2} + \frac{1}{2\sqrt{2}} & 0 \cr
        0 & \frac{1}{2\sqrt{2}} \cr
        0 & \frac{1}{2\sqrt{2}}\cr
        \frac{1}{2} - \frac{1}{2\sqrt{2}} & 0
    \end{pmatrix}, \qquad
    V_2 = \begin{pmatrix}
        0 & \frac{1}{2} - \frac{1}{2\sqrt{2}} \cr
        \frac{1}{2\sqrt{2}} & 0 \cr
        \frac{1}{2\sqrt{2}} & 0 \cr
        0 & \frac{1}{2} + \frac{1}{2\sqrt{2}}
    \end{pmatrix}.    
\end{equation}
These operators are related to those in Eq.~\!\eqref{eqn:phase_covariant} via a basis transformation. Specifically, by defining $T_1 = \frac{1}{\sqrt{2}}(K_1 - i K_2)$ and $T_2 = \frac{1}{\sqrt{2}}(K_2 - i K_1)$, the operators $V_i$ can be expressed as:
\begin{equation}
    V_i = (R_x \otimes R_X) T_i R_x^\dagger,\qquad   R_x = \frac{1}{\sqrt{2}} \begin{pmatrix} 1 & i \cr i & 1 \end{pmatrix}
\end{equation}
corresponds to a $\frac{\pi}{2}$-rotation around the $(\ket{+}\ket{-})$-axis of the Bloch sphere.

For states restricted to a great circle on the Block sphere, the SDP identifies a higher optimal fidelity compared to the universal case. For a $1 \to 2$ cloner, we obtain $F = \frac{1}{2} + \frac{\sqrt{2}}{4} \approx 0.8536$. The optimal fidelity for the $1 \to N$ case is given by \cite{Fan2001, Ariano2003, Bruss2000}:
\begin{equation}
    F^{opt}_{1 \mapsto N} = 
    \begin{cases}
        \dfrac{1}{2} + \dfrac{\sqrt{N(N+2)}}{4N}, & N \text{ is even}, \\[6pt]
        \dfrac{1}{2} + \dfrac{(N+1)}{4N}, & N \text{ is odd}.
    \end{cases}
\end{equation}
Table \ref{tab:covariant1toNcloning} in the appendix compares these analytical bounds with our numerical results, showing perfect agreement.

For the phase-covariant $1 \mapsto 3$ cloner, the reconstruction yields a single Kraus operator, which can be straightforwardly extended to a unitary matrix by augmenting the auxiliary space. This corresponds to an economical implementation of the cloning operation. The Kraus operator $K$ and the corresponding unitary matrix $U$ are given by:
\begin{equation}
    K = \begin{pmatrix}
        0 & 0 \cr
        a & 0 \cr
        a & 0 \cr
        0 & a \cr
        a & 0 \cr
        0 & a \cr
        0 & a \cr
        0 & 0
    \end{pmatrix},\qquad
    U = \begin{pmatrix}
        0 & b            & b            & c            & 0 & c            & c            & 0 \\
        a & \frac{2}{3}  & -\frac{1}{3} & d            & 0 & d            & d            & 0 \\
        a & -\frac{1}{3} & \frac{2}{3}  & d            & 0 & d            & d            & 0 \\
        0 & 0            & 0            & \frac{2}{3}  & a & -\frac{1}{3} & -\frac{1}{3} & 0 \\
        a & -\frac{1}{3} & -\frac{1}{3} & e            & 0 & e            & e            & 0 \\
        0 & 0            & 0            & -\frac{1}{3} & a & \frac{2}{3}  & -\frac{1}{3} & 0 \\
        0 & 0            & 0            & -\frac{1}{3} & a & -\frac{1}{3} & \frac{2}{3}  & 0 \\
        0 & 0            & 0            & 0            & 0 & 0            & 0            & 1
    \end{pmatrix},
\end{equation}
where 
\begin{equation}
    a = -\frac{\sqrt{3}}{6} + i \frac{1}{2},\quad b = \frac{\sqrt{3}}{6} + i \frac{1}{2},\quad c = -\frac{1}{6} + i \frac{\sqrt{3}}{6},\quad d = -\frac{\sqrt{3}}{18} - i \frac{1}{6},\quad e = \frac{\sqrt{3}}{9} + i \frac{1}{3}
\end{equation}
The cloning operation transforms the input state $\rho_\text{in}$ to the output state via $K \rho_\text{in} K^\dagger$ for the Kraus representation, or equivalently $U (\rho_\text{in} \otimes [(\ket{0}\bra{0})^{\otimes 2}]  ) U^\dagger$ for the unitary implementation.

\subsection{Cloning Two Pairs of Orthogonal States}
\begin{figure}[h]
    \centering
    \includegraphics[height=5cm]{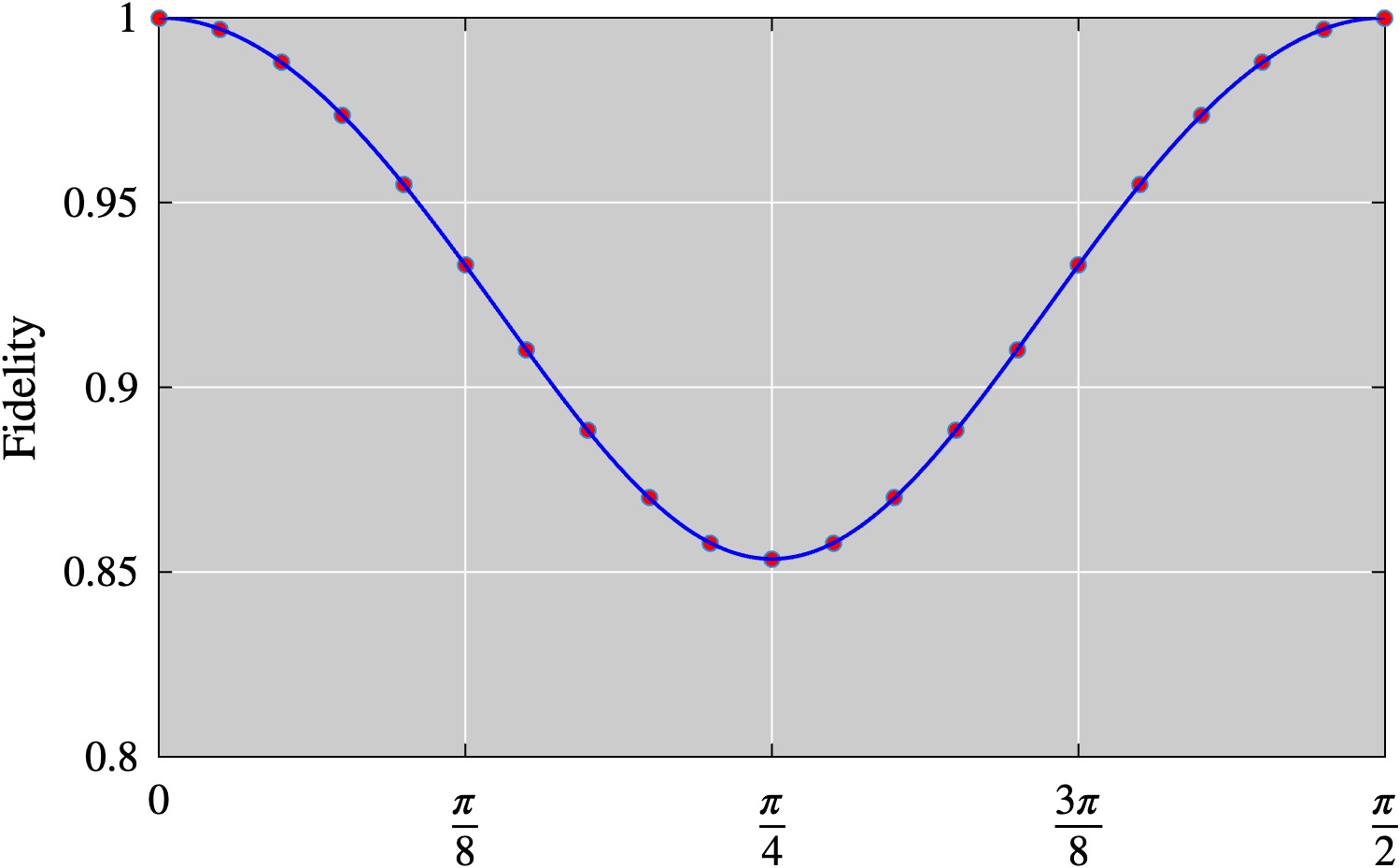}
    \caption{Numerically calculated fidelities (dots) compared to the analytical bound $F^{opt}(\theta)$ (solid blue line).}
    \label{FidelityTwoNonOrthogonalStates}
\end{figure}

Bruß and Macchiavello \cite{Bruss2001} investigated the optimal cloning of two pairs of orthogonal states and analytically derived the corresponding optimal fidelities . To reproduce their results, we consider the orthogonal pair $\{\ket{0}, \ket{1}\}$ and a second pair rotated by an angle $\theta$:
\begin{equation}
    \ket{0},\quad \ket{1}, \quad R(\theta)\ket{0}, \quad R(\theta)\ket{1} \quad\textrm{with}\quad 
    R(\theta) = \begin{pmatrix}
        \cos\theta &  -\sin\theta ) \\
        \sin\theta &   \cos\theta )
    \end{pmatrix}.
\end{equation}
In this case, we can construct $\Omega$ directly from these four states. 
By calculating the optimal fidelity as a function of the rotation angle $\theta$, we obtain the values shown in Figure \ref{FidelityTwoNonOrthogonalStates}. Our numerical findings perfectly recover the analytical result from \cite{Bruss2001}:
\begin{equation}
    F^{opt}(\theta) = \frac{1}{2} \left( 1 + \sqrt{ \sin^4 \theta + \cos^4\theta } \right).
\end{equation}

\subsection{Asymmetric Cloning}\label{sec:asymmetric}
In the asymmetric case, we relax the requirement of identical outputs to study the distribution of information between two parties \cite{Cerf2000c}. By introducing a trade-off parameter $\lambda \in [0, 1]$, we optimize the weighted objective $\lambda F_A + (1-\lambda) F_B$. The SDP maps the Pareto frontier of the cloning process, defining the boundary of physically allowed fidelity pairs $(F_A, F_B)$. For each point on this frontier, the framework provides a specific set of Kraus operators that achieve the optimal balance. For instance, when $\lambda$ is large, the extracted Kraus operators represent a channel that prioritizes the fidelity of subsystem $A$ while introducing more noise into subsystem $B$. This analysis is particularly relevant for quantifying the information gain of an eavesdropper in quantum cryptographic protocols. Note that for $\lambda = 0.5$, both clones yield the same symmetric fidelity value (e.g., $5/6 \approx 0.8333$ for the universal case).

\begin{figure}[h]
    \centering
    \includegraphics[height=3.5cm]{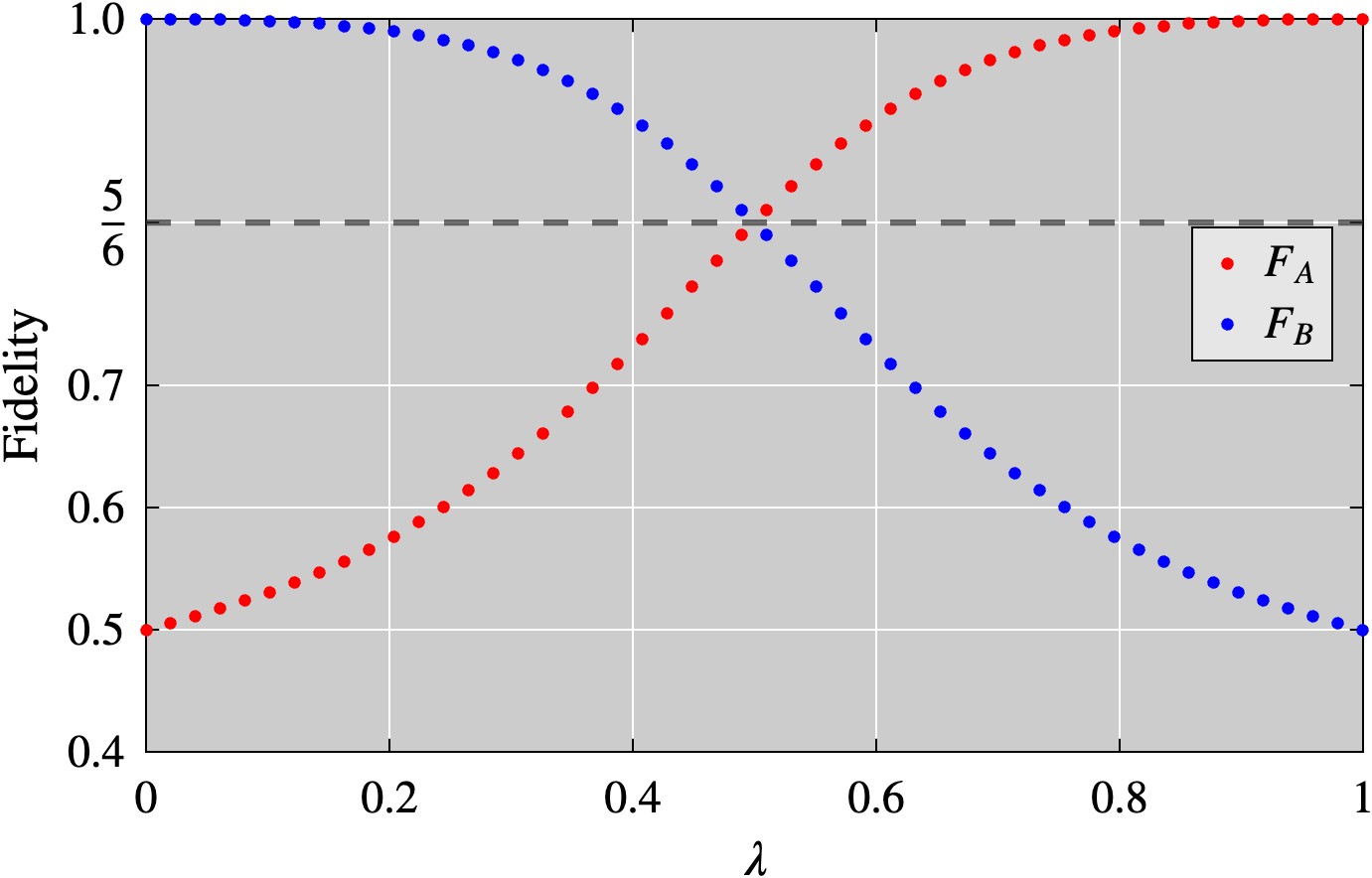}\hspace{5 mm}
    \includegraphics[height=3.5cm]{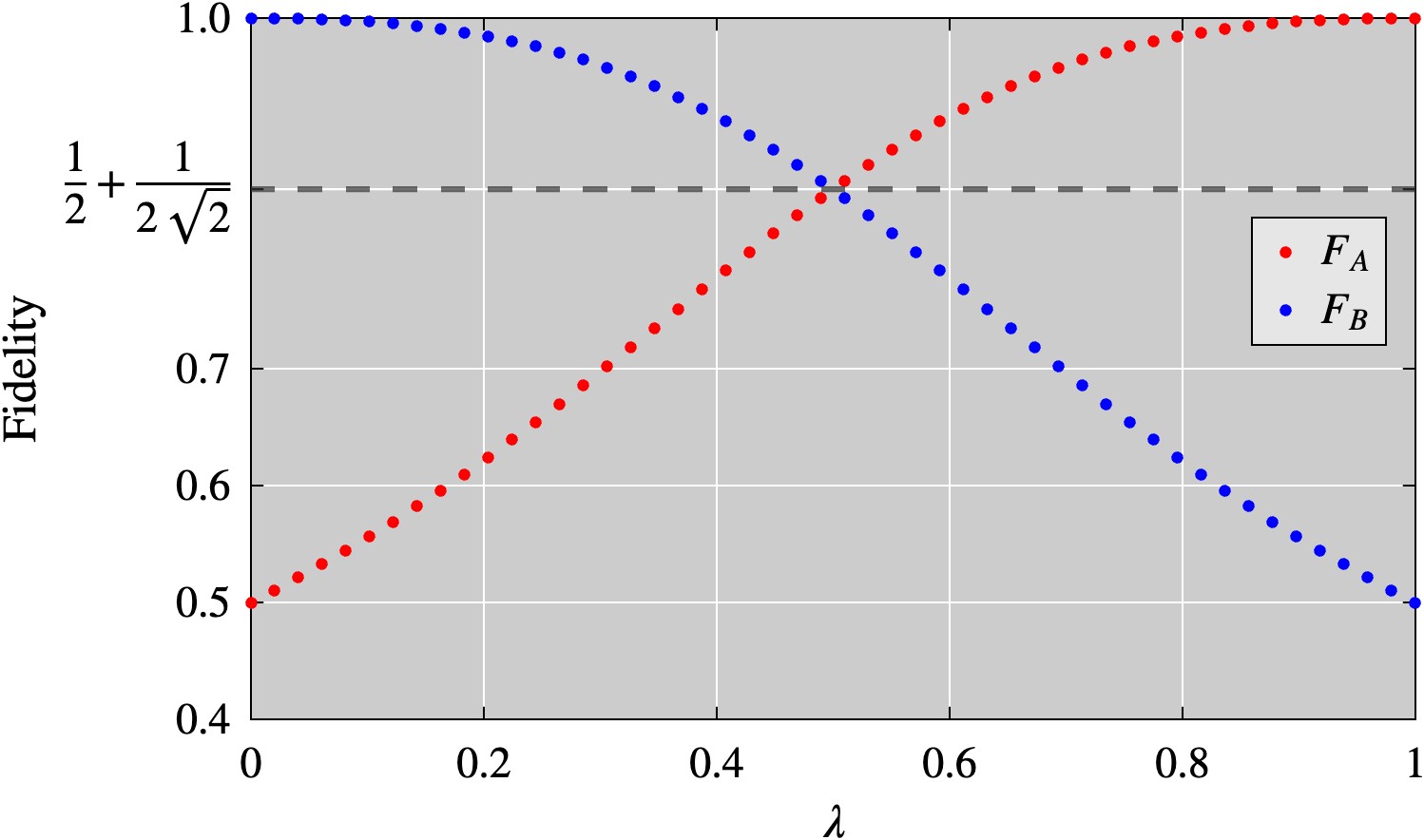}
    \caption{Numerically calculates fidelities for asymmetric cloning. Left for univeral and right for covariant cloning. For $\lambda = 0.5$ we find the values for the symmetric cloning.}
    \label{AsymmetricFidelities}
\end{figure}

To implement this numerically, we define two target operators:
\begin{eqnarray}
    \Omega_A &=& \frac{1}{|\mathcal{S}|} \sum_{|\psi_k\rangle \in \mathcal{S}} \left( \ket{\psi^*_k} \bra{\psi^*_k}\right)_\text{in} \otimes \left( \ket{\psi_k} \bra{\psi_k} \right)_{{\text{out}}_A} \otimes \mathbb{1}_2 \\
    \Omega_B &=& \frac{1}{|\mathcal{S}|} \sum_{|\psi_k\rangle \in \mathcal{S}} \left( \ket{\psi^*_k} \bra{\psi^*_k}\right)_\text{in} \otimes \mathbb{1}_2 \otimes \left( \ket{\psi_k} \bra{\psi_k} \right)_{{\text{out}}_B} 
\end{eqnarray}
Since there is only one input qubit, the known SIC states can be used as sample sets.
The SDP then maximizes the weighted objective:
\begin{equation}
    J \cdot \Omega(\lambda) = J \cdot \left( \lambda \cdot \Omega_A + (1 - \lambda)\cdot \Omega_B \right) 
\end{equation}
Figure \ref{AsymmetricFidelities} displays the fidelities of the two output qubits. The left plot shows the results for asymmetric universal cloning, while the right plot shows the results for asymmetric covariant cloning. As expected, at $\lambda = 0.5$, the results converge to the previously discussed symmetric fidelity values.

\subsection{Entanglement Cloning} \label{sec:entanglementConling}
We can extend our framework to study the cloning of pure entangled states, which represents a special case of $2 \to 4$ cloning. Here, we consider a four-dimensional input system and a 16-dimensional output system. The symmetry constraint is applied to each of the two four-dimensional bipartite subsystems as a whole. For cloning general maximally entangled two-qubit states, our numerical framework reproduces the fidelity value of $F = (5 + \sqrt{13})/12 \approx 0.7171$, as reported in \cite{Lamoureux2004}.

As a further application, we analyze the cloning of maximally entangled states with real amplitudes:
\begin{equation}
    \ket{\Psi} = c_{00} \ket{00} + c_{01} \ket{01} + c_{10} \ket{10} + c_{11} \ket{11} 
\end{equation}
where $c_{00},c_{01},c_{10},c_{11} \in \R$, satisfy $c_{00}^2 + c_{01}^2 + c_{10}^2 + c_{11}^2 = 1$ and the maximal entanglement condition  $|c_{00}c_{11} - c_{01}c_{10}| = \frac{1}{2}$.
In this case, we can choose the four Bell states $\ket{\Psi^\pm}, \ket{\Phi^\pm}$ and the derived states through the application of $H \otimes \mathbb{1}$: 
\begin{equation}
    \begin{split}
        \ket{\Psi_1} &= \frac{1}{2}(-\ket{00} + \ket{01} + \ket{10} + \ket{11}) \\
        \ket{\Psi_2} &= \frac{1}{2}(\ket{00} - \ket{01} + \ket{10} + \ket{11}) \\ 
        \ket{\Psi_3} &= \frac{1}{2}(\ket{00} + \ket{01} - \ket{10} + \ket{11}) \\
        \ket{\Psi_4} &= \frac{1}{2}(\ket{00} + \ket{01} + \ket{10} - \ket{11})
    \end{split}
\end{equation}
for the construction of $\Omega$.

Our results show that the cloning of an arbitrary maximally entangled two-qubit state with real coefficients achieves an optimal fidelity of $F_{\text{opt}} = \frac{1}{2} + \frac{1}{2\sqrt{2}} \approx 0.8536$. The resulting clones exhibit a purity of $3/4$ and a concurrence of $1/\sqrt{2}$. The algebraic forms of the corresponding Kraus operators, derived from our numerical results, are provided in Appendix \ref{app:entanglement_kraus}.

\subsection{Summary of Numerical Validation}
Across all evaluated scenarios, the duality gap consistently remained below $10^{-7}$, and the completeness relation (trace-preserving condition) $\sum_i K_i^\dagger K_i = \mathbb{1}_{\text{in}}$ was satisfied with a numerical error of less than $10^{-8}$. These metrics provide a rigorous certification that the calculated fidelities and the extracted Kraus operators represent the globally optimal quantum channels for the respective cloning tasks. The excellent agreement with analytical values reported in the literature further confirms the robustness and precision of our numerical framework.

\section{Cloning attack on the BB84 protocol, taking noise into account}\label{sec:BB84Attack}
\begin{figure}[h]
    \centering
    \includegraphics[height=7cm]{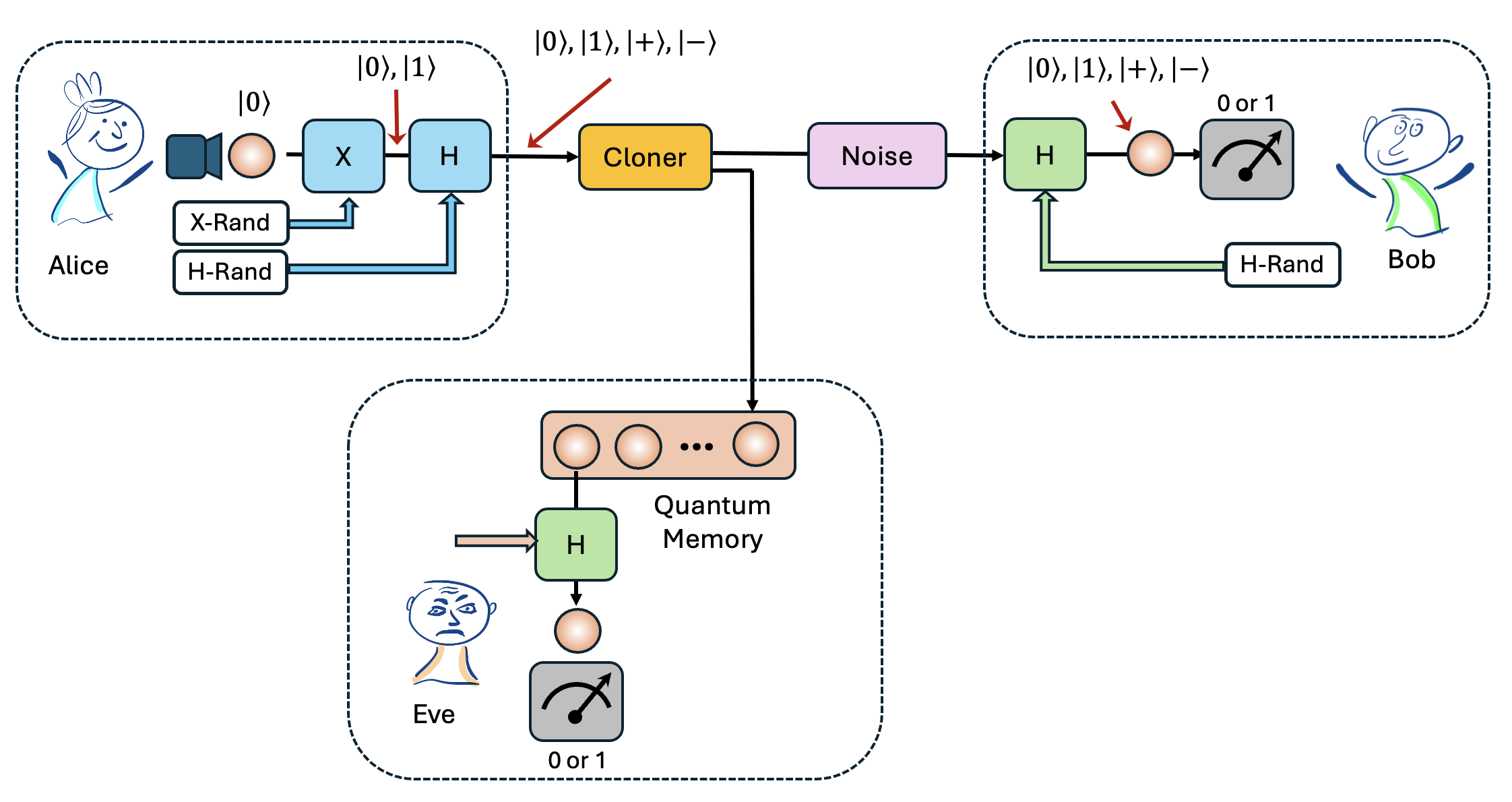}
    \caption{BB84 Cloning Attack: Alice sends random qubits in the states $\ket{0}, \ket{1}, \ket{+}$, or $\ket{-}$. Eve intercepts and clones the qubits, storing her copy in a quantum memory. Once Alice and Bob have sifted their key, Eve uses the basis information to measure her clones and reconstruct the key.}
    \label{BB84Attack}
\end{figure}

Describing quantum processes using Kraus operators allows for the analysis of complex scenarios with significant ease. As a demonstration, we analyze a cloning attack on the BB84 protocol \cite{BB84}, accounting for a depolarizing disturbance. This example extends the analysis presented in \cite{Pigott2025}. We assume that depolarization occurs during the transmission to Bob after the cloning attack has taken place (cf. Fig. \ref{BB84Attack}). Note that the disturbance itself is represented as a quantum channel. For a comprehensive review of practical security results, see also \cite{Scarani2009}.

We assume the attacker can store her cloned qubits. This allows her, after eavesdropping on the classical basis exchange, to perform the appropriate measurement (e.g., applying a Hadamard operation if the diagonal basis was used). Furthermore, we assume that the eavesdropper (Eve) uses an asymmetric cloner (parameter $\lambda$) and that the noise intensity can be varied via a parameter $\mu$.

To analyze this scenario, we concatenate two quantum channels. Let $\rho_{in}$ be the input to the cloner; the output of the cloning stage is:
\begin{equation}
    {\cal E}_{clone}(\rho_{in}) = \sum_k K_k \rho_{in} K_k^\dagger
\end{equation} 
In our model, depolarization affects only Bob's qubit, while Eve's clone remains unaffected. The output of the serially concatenated channel is given by \cite{Wilde_2017}:
\begin{equation}
    \begin{split}
        ({\cal N} \circ {\cal E})(\rho_{in}) &= \sum_l (N_l \otimes \mathbb{1}_2 ) \left( {\cal E}_{clone}(\rho_{in}) \right)  (N_l^\dagger \otimes \mathbb{1}_2) \\
        &= \sum_{l,k} (N_l \otimes \mathbb{1}_2 ) \left( K_k \rho_{in} K_k^\dagger \right)  (N_l^\dagger \otimes \mathbb{1}_2),
    \end{split}
\end{equation} 
where $\mathcal{N} = \sum_{l=0}^3 {\cal N}_l, (0 \leq \mu \leq \frac{3}{4})$ is the standard depolarizing channel with operators:
\begin{equation}
    N_0 = \sqrt{1-\mu} \hspace{1 mm} \mathlarger{ \mathbb{1} }, \quad
    N_1 = \sqrt{ \frac{\mu}{3} } X, \quad
    N_2 = \sqrt{ \frac{\mu}{3} } Y, \quad
    N_3 = \sqrt{ \frac{\mu}{3} } Z.
\end{equation}
where $X,Y$ and $Z$ are the Pauli matrices.

\begin{figure}[h]
    \centering
    \includegraphics[height=3.5 cm]{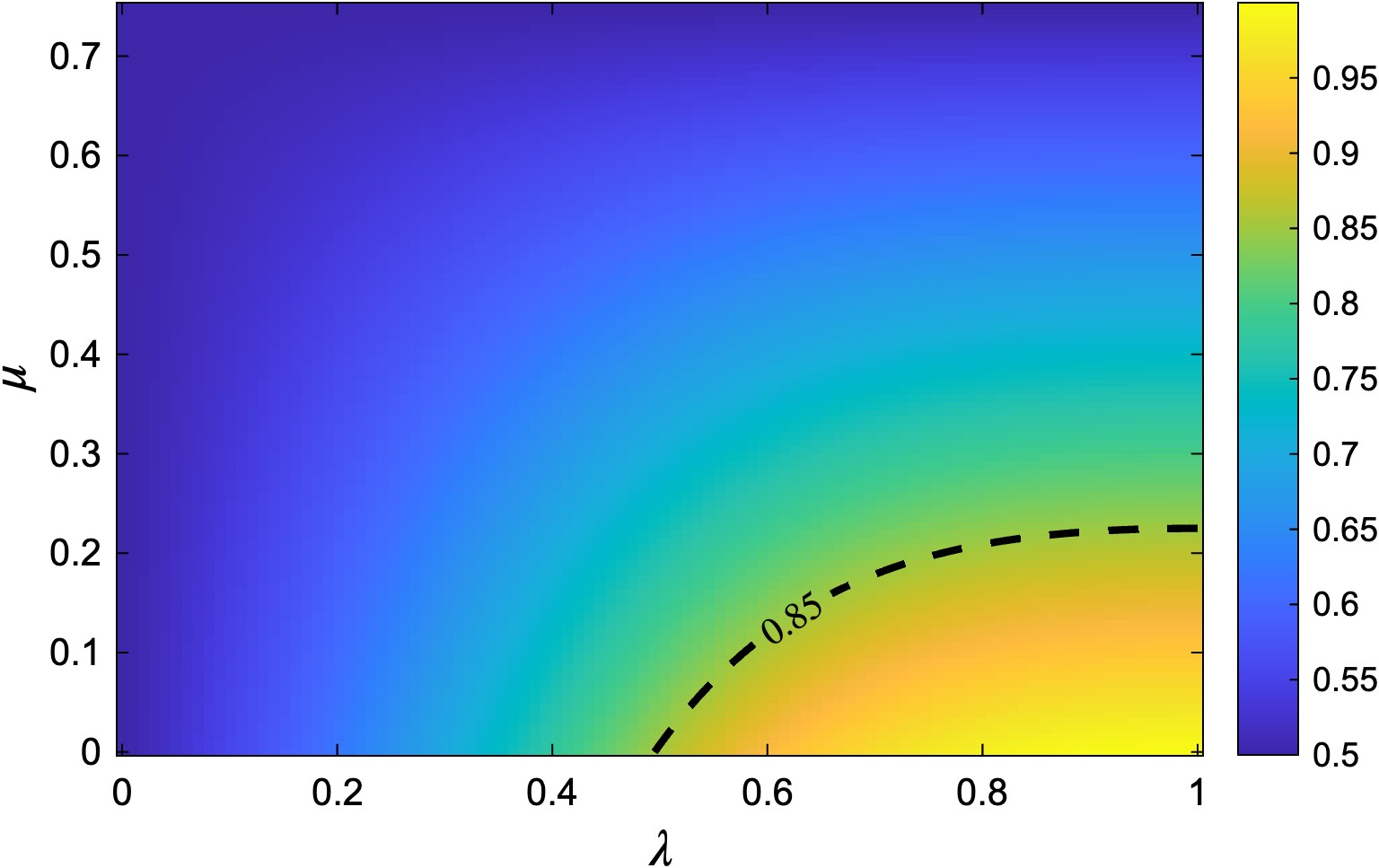}
    \caption{Fidelity of Bob's qubits as a function of asymmetry $\lambda$ and depolarization $\mu$ for the four-state protocol. The dashed line indicates the 15\% QBER threshold ($F_{\text{Bob}} = 0.85$). Bob's and Eve's qubits are in this scenario not entangled.}.
    \label{BB84Fidelity4States}
\end{figure}

\begin{figure}[h]
    \centering
    \includegraphics[height=3.5 cm]{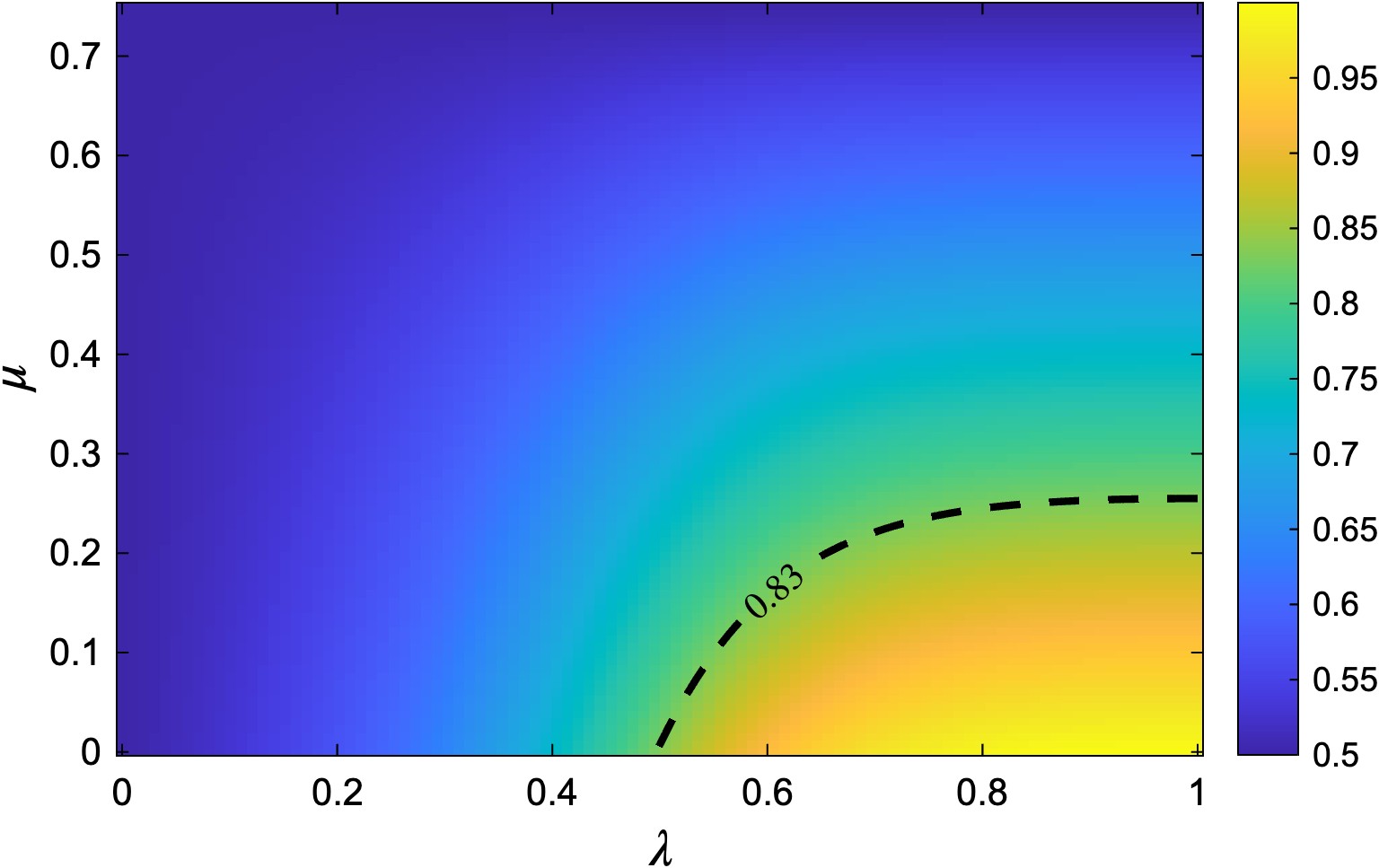}\hspace{3 mm}
    \includegraphics[height=3.5 cm]{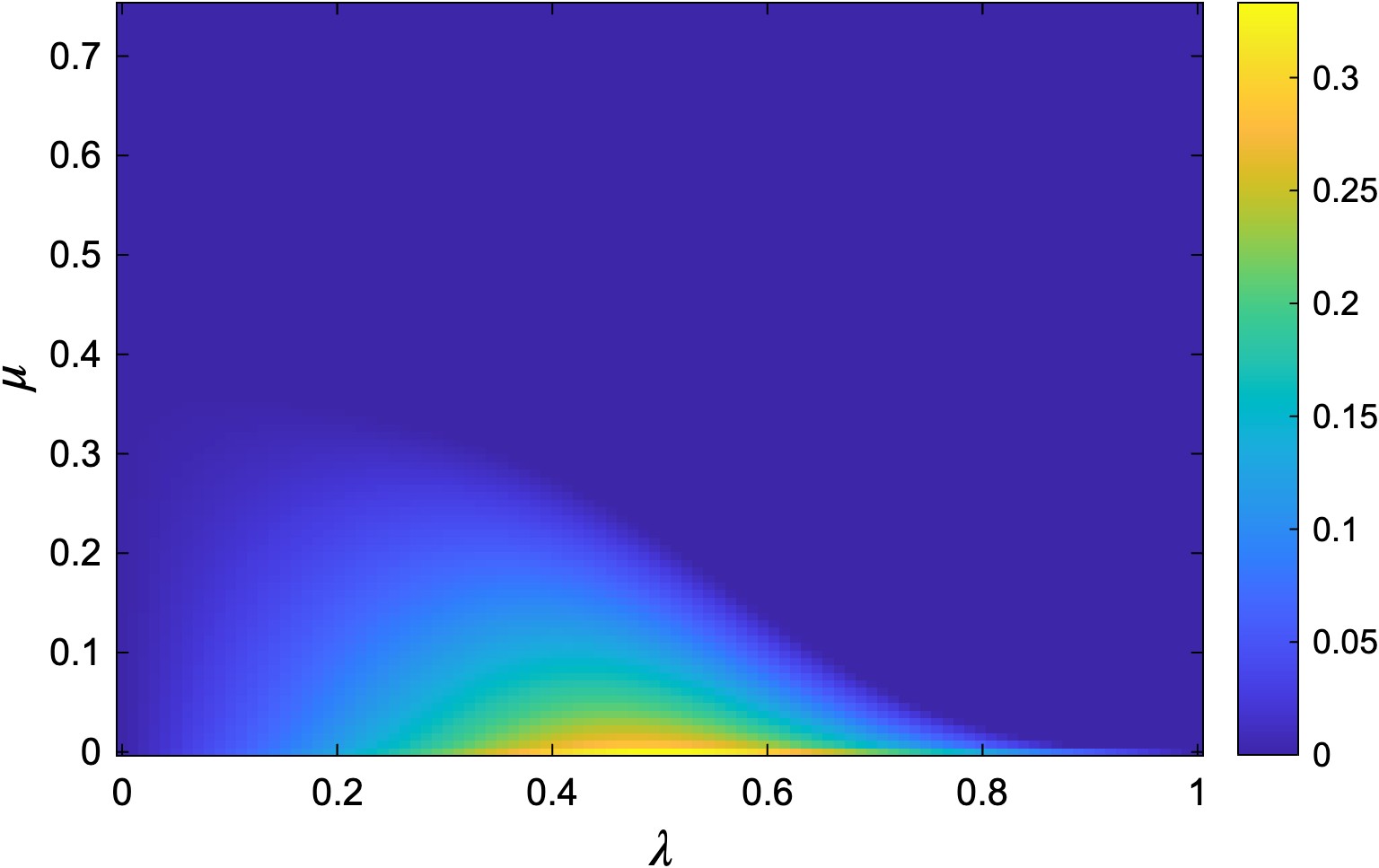}
    \caption{Left: Bob's fidelity for the six-state protocol. The dashed line corresponds to the security limit $F_{\text{Bob}} \approx 0.832$ (corresponding to $QBER \approx 12.6\%$). Right: Concurrence between Bob's and Eve's qubits, showing the presence of entanglement.}
    \label{BB84Fidelity6States}
\end{figure}

Figures \ref{BB84Fidelity4States} and \ref{BB84Fidelity6States} show Bob’s fidelity as a function of the cloning asymmetry $\lambda$ and the depolarization strength $\mu$. The results depend on the state set used for transmission, specifically whether the four-state (BB84) or the six-state protocol is employed. We consider here an individual attack, where Eve interacts with each qubit independently. Fuchs et al. showed that Alice and Bob can distill a secure key provided the Quantum Bit Error Rate (QBER) is below approximately 15\% \cite{Fuchs1997}. For collective attacks, where Eve may measure her stored qubits collectively, the security threshold is 11\% \cite{Pirandola2008}. For the six-state variant \cite{Bruss1998SixState}, the threshold is higher, with Lo reporting a limit of 12.6\% \cite{Le2001}.

The QBER can be derived directly from the fidelity $F_{\text{Bob}} = \bra{\psi} \rho_B \ket{\psi}$. For the depolarizing channel, the relations are:
\begin{eqnarray}
    QBER &=& 1 - F_{Bob}, \qquad \textrm{(classical BB84, four state protocol)} \\
    QBER &=& \frac{3}{4}(1 - F_{Bob}), \quad \textrm{(six-state protocol)} 
\end{eqnarray}

Since the focus of this work is the application of the SDP framework rather than the derivation of new security bounds, we provide these results as a validation of the framework's capability to handle multi-stage quantum processes.

\section{Discussion and Extension} \label{sec:discussion}

The results presented in the previous section demonstrate that the SDP-based framework is a powerful and versatile tool for analyzing quantum cloning. In this section, we discuss the physical implications of our findings and outline possible extensions to more complex quantum information tasks.

\subsection{Scalability and Computational Limits}
While the SDP approach provides global optima, it is subject to the curse of dimensionality. The dimension of the Choi matrix for an $M \to N$ cloning process scales as $d^{M+N}$, where $d$ is the dimension of the individual subsystems. Consequently, small qubit systems such as $1 \to 2$ or $1 \to 3$ can be solved within seconds on standard hardware. For larger $N$, however, the number of variables in the SDP increases exponentially, leading to a significant rise in runtime and memory requirements.

To overcome this limitation in future iterations of this framework, a highly 
promising avenue is the systematic integration of group-theoretic symmetry 
reduction. Because optimal quantum cloning channels inherently exhibit invariance 
under specific group actions -- such as the symmetric group $S_M$ for permutation 
invariance of the output copies, or the unitary group $SU(2)$ for universal 
cloning -- the underlying SDP can be substantially simplified. 

By exploiting the Schur--Weyl duality, the large, full-rank Choi matrix can be 
projected onto the corresponding commutant algebra (the centralizer). This allows 
for an explicit block-diagonalization of the optimization variables into 
irreducible representations associated with specific Young diagrams. 
Implementing this representation-theoretic reduction will transform the computational 
complexity from exponential to polynomial scaling. Consequently, this symmetry-aware 
extension will enable the numerical certification and Kraus-operator extraction 
of higher-order, multi-qubit cloning channels that are currently intractable 
within the unreduced state space.

\subsection{Extension to Noisy Cloning and Broadcasting}
An interesting path for future research is the inclusion of pre-existing noise. Future work should examine the extent to which this framework can be adapted to scenarios where the input states are already mixed or where the cloning channel itself is subject to additional constraints, such as Local Operations and Classical Communication (LOCC). Furthermore, the framework could be extended from cloning (pure-to-pure) to quantum broadcasting (mixed-to-mixed) by adjusting the target operator $\Omega$ to reflect the fidelity between mixed density operators.

\subsection{Economical Cloning}
Economical cloners are characterized by their ability to be implemented via unitary operations without the need for an additional ancilla system (other than the target qubits). It is known that economical cloners exist for specific scenarios, such as the phase-covariant $1 \to 2$ cloner \cite{Buscemi2005} or the $1 \mapsto 3$ cloner given in section label{sec:covariant}. A compelling open question is how to enforce the "economical" constraint within an SDP framework. Since the Choi matrix of a unitary process (or a pure state transformation) possesses only one non-zero eigenvalue, this would require rank-constraints, which typically lead to non-convex optimization problems (e.g., SDPs with rank-1 constraints).

Because economic cloning protocols strictly require 
the output copies to be symmetric under permutation, the entire target space 
can be restricted to the symmetric subspace. 

By applying representation-theoretic constraints, the dimension of the output 
space for $M$ qubits drops from $2^M$ to just $M+1$. Integrating this 
restriction into our SDP formulation will allow the framework to effortlessly 
optimize and extract Kraus operators for large-scale economic cloning networks, 
providing a powerful tool for analyzing resource-efficient eavesdropping strategies 
in quantum cryptography.

\subsection{Application to Quantum Cryptography}
The ability to compute the Pareto boundary for asymmetric cloning has direct implications for the security analysis of Quantum Key Distribution (QKD). In a cloning-based attack, an eavesdropper (Eve) acts as an asymmetric cloner, attempting to maximize her information gain ($F_B$) while minimizing the disturbance caused to the legitimate receiver ($F_A$). Our framework allows for a precise calculation of the maximum information leakage even in non-standard protocols where the signal states do not follow the typical symmetries of the BB84 or six-state protocols.

\section{Conclusion} \label{sec:conclusion}
In this work, we have developed and demonstrated a robust computational framework for the systematic analysis of quantum cloning tasks. The core contribution of this study is the integration of Semidefinite Programming (SDP) with the Choi-Jamiołkowski isomorphism into a unified pipeline. This approach automates the transition from a formal physical problem definition to a concrete experimental blueprint.

By employing this framework, we have shown that a wide range of cloning scenarios — from universal and phase-covariant to asymmetric and entanglement-based tasks — can be solved with high numerical precision. Our results for established cloning models serve as a rigorous benchmark; the perfect agreement between the numerical outputs and known analytical bounds validates the reliability of the underlying algorithm. Beyond mere fidelity calculations, the framework's primary strength lies in its versatility: it handles arbitrary input state distributions and enforces structural constraints, such as symmetry requirements between output registers, within a single optimization step.

A significant utility of the presented pipeline is the automated extraction of Kraus operators from the optimized Choi matrices. This feature provides an operational description of the cloning process, enabling the direct design of quantum circuits and the characterization of noise processes in protocols like BB84.

In conclusion, the developed computational framework provides a versatile and extensible platform for exploring the limits of quantum information processing. It represents a powerful tool for researchers to efficiently navigate the landscape of optimal quantum transformations, especially in scenarios where analytical solutions are elusive or impossible to derive.

\subsection*{Code Availability}
The numerical pipeline developed in this work is publicly available at \url{https://github.com/jhettel/quantum-cloning-sdp}. This includes the SDP implementation and the benchmarks presented in the appendices.

\bibliography{bibliography}

\clearpage
\begin{appendices}

\section{Choi-Jamiołkowski Inner Product Identity}    \label{app:choi_derivation}
In this section, we derive the identity used to link the channel action to the Choi matrix:
\begin{equation}
    \langle \psi | {\cal E}(\ket{\phi}\bra{\phi}) | \psi \rangle = \langle \phi^* \otimes \psi | J({\cal E}) | \phi^* \otimes \psi \rangle
\end{equation}
Starting with the right-hand side (RHS) and substituting the definition of the Choi matrix $J({\cal E})$ into the inner product:
\begin{equation}
    \text{RHS} = \sum_{i,j} \langle \phi^* \otimes \psi | \left( |i\rangle\langle j| \otimes {\cal E}(|i\rangle\langle j|) \right) | \phi^* \otimes \psi \rangle
\end{equation}
Using the property of tensor products $(A \otimes B)(u \otimes v) = (Au) \otimes (Bv)$, we separate the input and output terms:
\begin{equation}
    \text{RHS} = \sum_{i,j} \langle \phi^* | i \rangle \langle j | \phi^* \rangle \cdot \langle \psi | {\cal E}(|i\rangle\langle j|) | \psi \rangle
\end{equation}
Given the basis expansion $|\phi\rangle = \sum_k c_k |k\rangle$, its complex conjugate in the same basis is $|\phi^*\rangle = \sum_k c_k^* |k\rangle$. The inner products satisfy:
\begin{equation}
    \langle \phi^* | i \rangle = (c_i^*)^* = c_i = \langle i | \phi \rangle
\end{equation}
Similarly, for the adjoint:
\begin{equation}
    \langle j | \phi^* \rangle = c_j^* = \langle \phi | j \rangle
\end{equation}
Substituting these coefficients into the sum, we obtain:
\begin{equation}
    \text{RHS} = \sum_{i,j} \langle i | \phi \rangle \langle \phi | j \rangle \cdot \langle \psi | {\cal E}(|i\rangle\langle j|) | \psi \rangle
\end{equation}
Due to the linearity of the map $\mathcal{E}$ and the inner product, the summation and the scalar coefficients can be moved inside the argument of the channel:
\begin{equation}
    \text{RHS} = \langle \psi | {\cal E} \left( \sum_{i,j} \langle i | \phi \rangle |i\rangle\langle j| \langle \phi | j \rangle \right) | \psi \rangle
\end{equation}
Recognizing the resolution of the identity or simply the outer product expansion, we see that $\sum_i |i\rangle \langle i | \phi \rangle = |\phi\rangle$ and $\sum_j \langle \phi | j \rangle \langle j| = \langle \phi |$. This directly leads to:
\begin{equation}
    \text{RHS} = \langle \psi | {\cal E} ( |\phi\rangle\langle\phi| ) | \psi \rangle = \text{LHS}
\end{equation}

\section{Numerical Results and Certification}\label{app:NumericalResults}
This section provides comprehensive numerical results obtained using the SDP framework described in Section \ref{sec:results}. For many of the benchmarks, the numerical values can be identified with their exact algebraic forms and verified through direct substitution into the theoretical fidelity expressions. The data provided, including the optimized Kraus operators, are intended to facilitate the reconstruction and verification of these quantum channels in standard simulation environments such as \textit{QuTiP} or \textit{Qiskit}.

\subsection{Sampling Set}
Table \ref{tab:symmetric1to2cloning} presents the results of the primal and dual SDP for a universal symmetric $1 \to 2$ cloner, depending on the size and nature of the sampling set $\cal S$ used to construct $\Omega$ (see Sec. \ref{sec:sampling_set}).

In addition to the four informationally complete states (SIC-POVM) (see Eqn. \ref{eqn:sic-povm}), we examined in addition a fixed set of six states (the eigenbases of the Pauli matrices):
\begin{equation}\label{eqn:pauli_set}
    \ket{0}, \ket{1}, \ket{+} = \frac{\left(\ket{0} + \ket{1}\right)}{\sqrt{2}} , \ket{-} = \frac{\left(\ket{0} - \ket{1}\right)}{\sqrt{2}} , \ket{i} = \frac{\left(\ket{0} + i \ket{1}\right)}{\sqrt{2}} , \ket{-i} = \frac{\left(\ket{0} - i \ket{1}\right)}{\sqrt{2}} 
\end{equation}
For other configurations, we distributed the qubits across the Bloch sphere using a Spherical Fibonacci Lattice. This method ensures a near-uniform and isotropic distribution of $N$ points by maintaining nearly constant Voronoi cell areas across the surface, while also providing a reproducible point set.

\begin{table}[h]
    \centering
    \caption{Comparison of different $\Omega$ sampling strategies for the universal symmetric $1 \to 2$ cloner (theoretical fidelity $F = 5/6 \approx 0.8333\bar{3}$). The table shows the primal and dual objective values and their respective gaps to the analytical bound. Note that for smaller $N$, the numerical optima may slightly exceed the theoretical limit due to the restricted sampling set.}
    \label{tab:symmetric1to2cloning}
    \begin{tabular}{lcccc}
        \toprule
        $|\cal S|$ & Primal & Primal Gap & Dual & Dual Gap  \\
        \midrule
        4 (SIC) & 0.8333333328 &  4.973e-10 & 0.8333333354 & 2.040e-09  \\[1ex]
        6 (fix) & 0.8333333328 &  4.973e-10 & 0.8333333354 & 2.040e-09  \\[1ex] \hline \\ 
        10      & 0.8358345438 & -2.501e-03 & 0.8358345467 & 2.501e-03  \\[1ex]
        100     & 0.8333473964 & -1.406e-05 & 0.8333473983 & 1.406e-05  \\[1ex]
        1000    & 0.8333334456 & -1.123e-07 & 0.8333334481 & 1.148e-07  \\[1ex]
        10000   & 0.8333333339 & -6.154e-10 & 0.8333333365 & 3.153e-09  \\[1ex]
        100000  & 0.8333333328 &  4.862e-10 & 0.8333333354 & 2.051e-09  \\[1ex]
        \bottomrule
    \end{tabular}
\end{table}.

\begin{figure}[h]
    \centering
    \includegraphics[height=6cm]{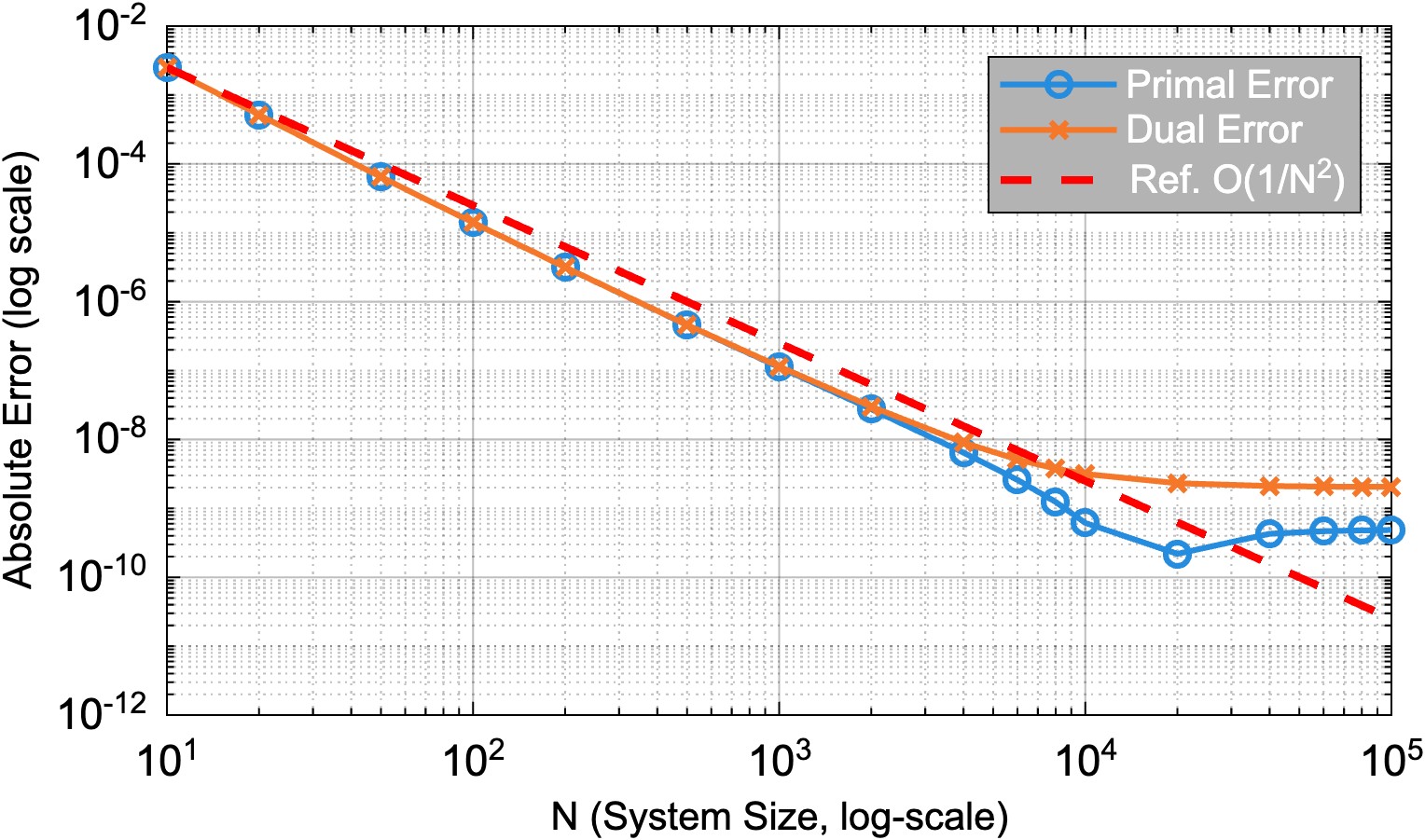}
    \caption{Convergence of the numerical error for primal and dual solutions ($|F_{opt} - primal|$ and $|F_{opt} - dual|$) as a function of the sampling set size $N$.}
    \label{Convergence_Analysis}
\end{figure}

The calculations demonstrate that numerical accuracy is highly dependent on the choice of the sample set $|\mathcal{S}|$. The highest precision is achieved using Symmetric Informationally Complete (SIC) states. These results are only matched by uniform distributions when the number of points is very large ($N \ge 10^5$). For uniform sampling, we observe that the error of the calculated fidelity converges to the theoretical value with a rate of approximately $\mathcal{O}(1/N^2)$ (see Fig. \ref{Convergence_Analysis}).
In each case, two Kraus operators were extracted from the resulting Choi matrix. For all pairs, the completeness relation was satisfied with a tolerance on the order of $10^{-9}$.

\begin{figure}[h]
    \centering
    \includegraphics[height=5cm]{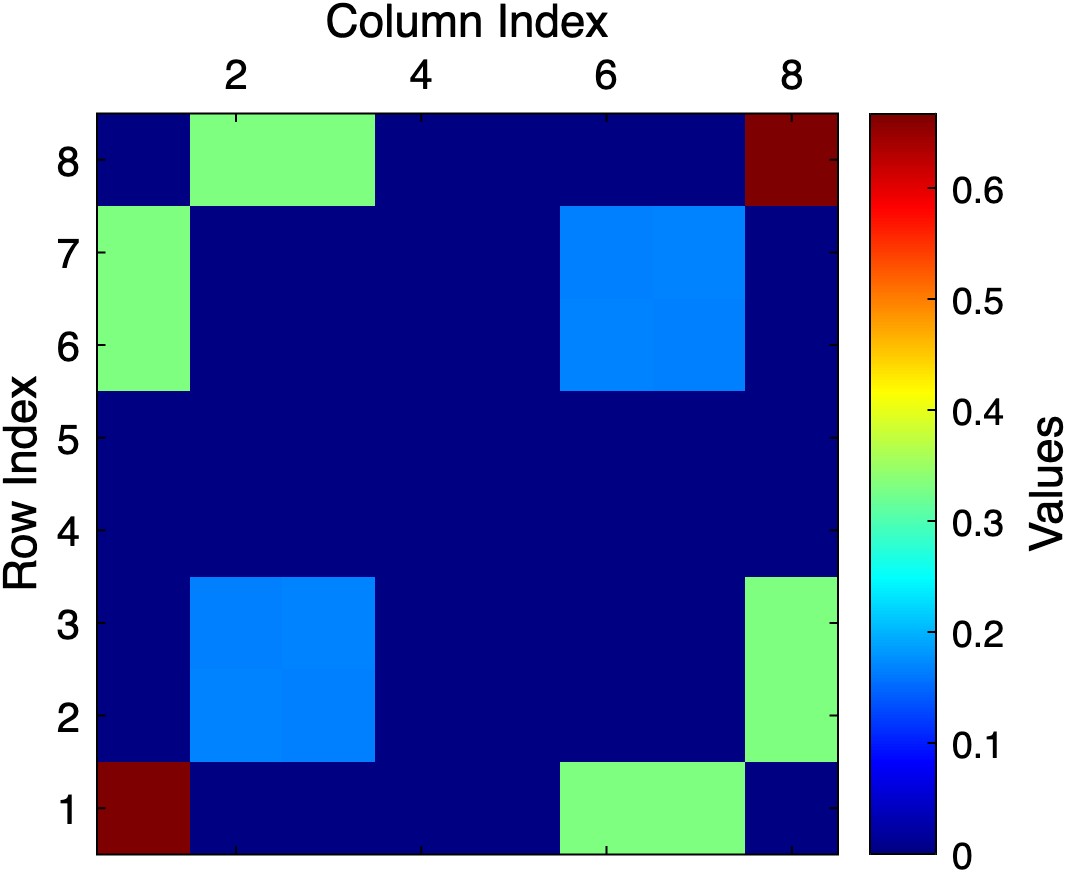}\hspace{3 mm}
    \includegraphics[height=4.5cm]{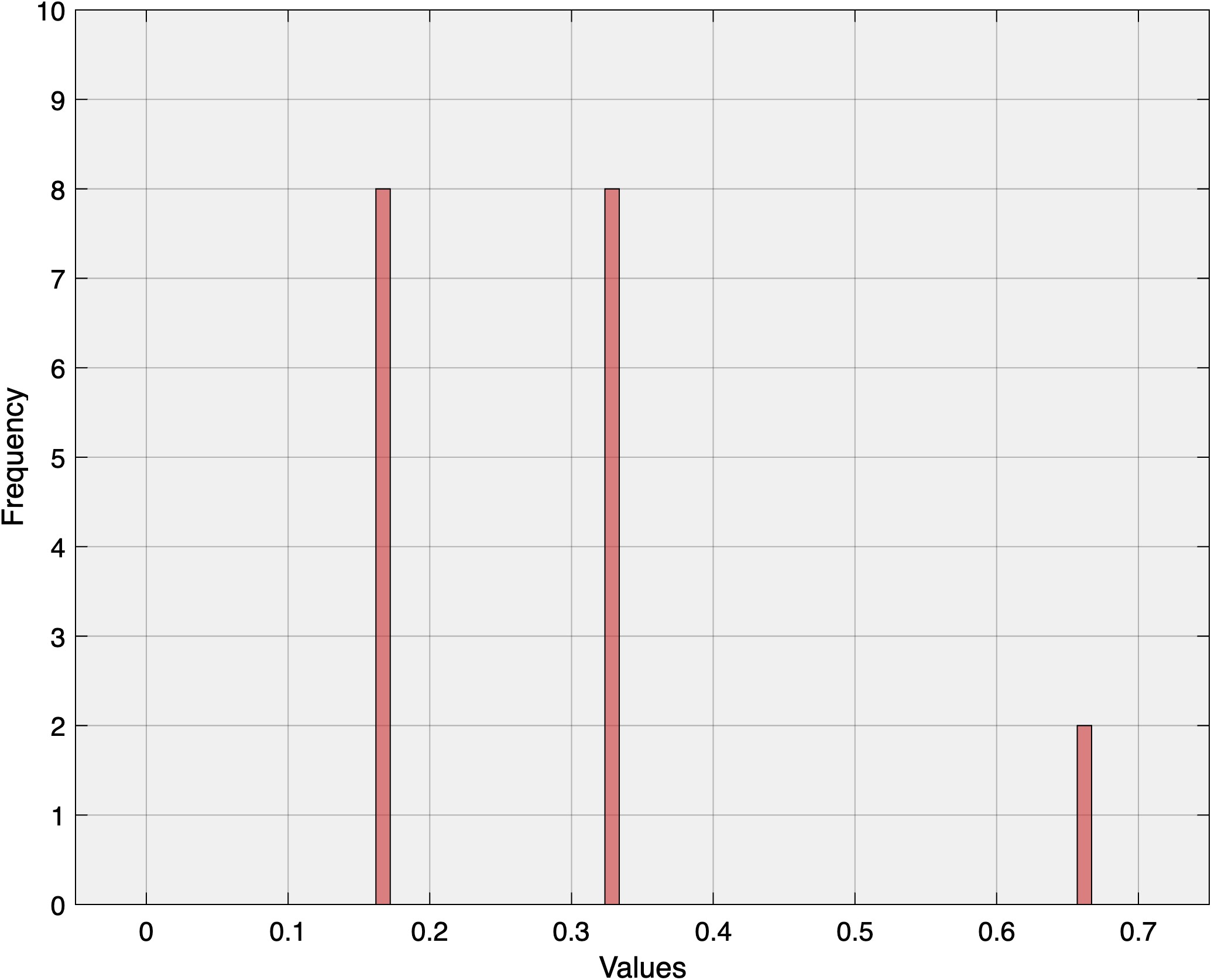}
    \caption{Structure of the Choi matrix for the $1 \mapsto 2$ cloner (left) and the frequencies of the non-zero entries (right) where $\mathcal{S}$ is constructed from SIC-POVM states.}
    \label{SIC_SamplingSet}
\end{figure}

An examination of the corresponding $8\times 8$ Choi matrices reveals that their symmetric structure emerges even for small sampling set sizes, albeit with significant noise in the form of numerous near-zero entries. Figure \ref{SIC_SamplingSet} presents a visualization of the Choi matrix (left panel) alongside the distribution of nonzero entries (right panel), where values below $10^{-9}$ are treated as numerical zeros. This example corresponds to the $1 \mapsto 2$ cloner where the sampling set $\mathcal{S}$ is constructed from SIC-POVM states (Eq.~\ref{eqn:sic-povm}).

\begin{figure}[h]
    \centering
    \includegraphics[height=4.5cm]{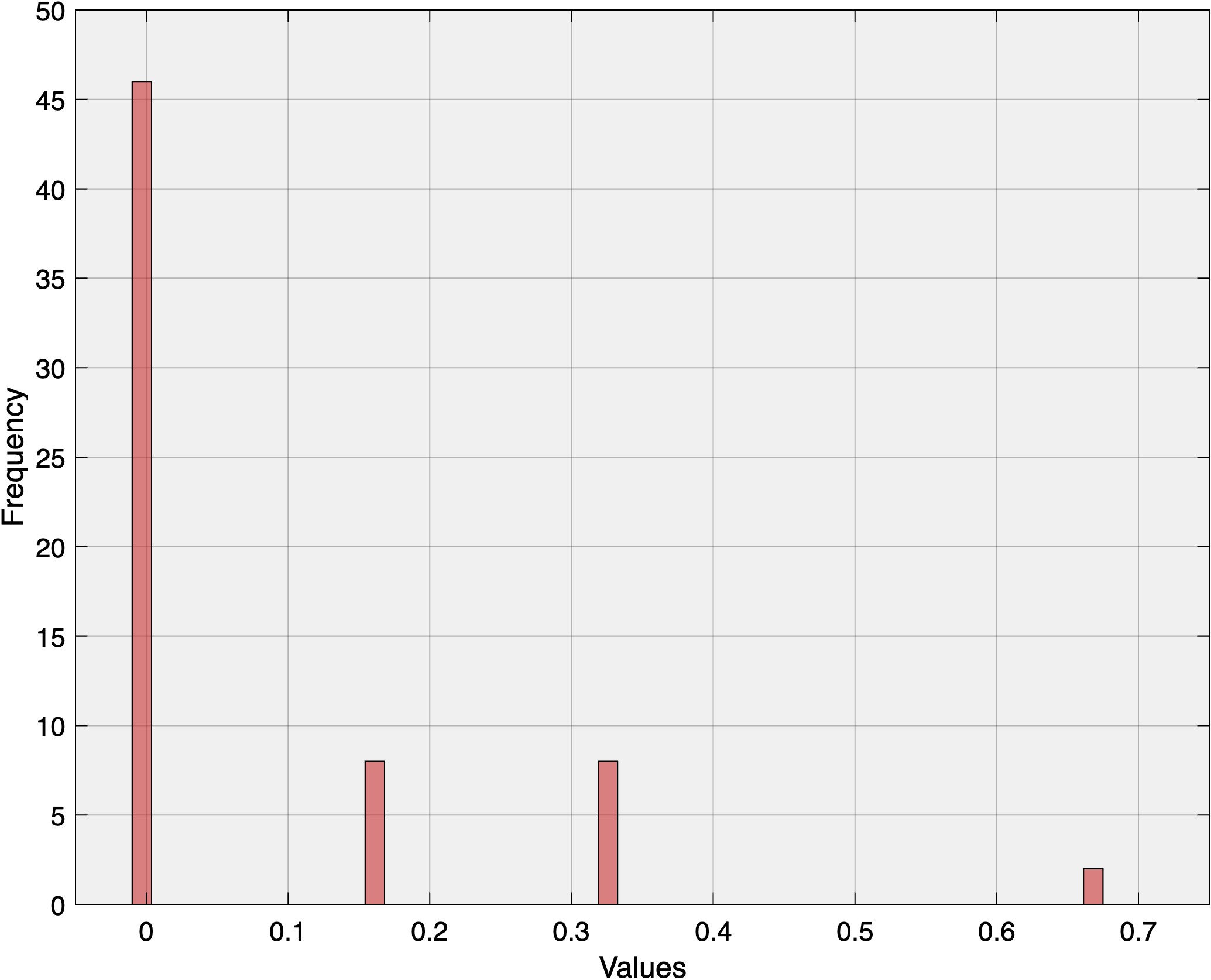}\hspace{6 mm}
    \includegraphics[height=4.5cm]{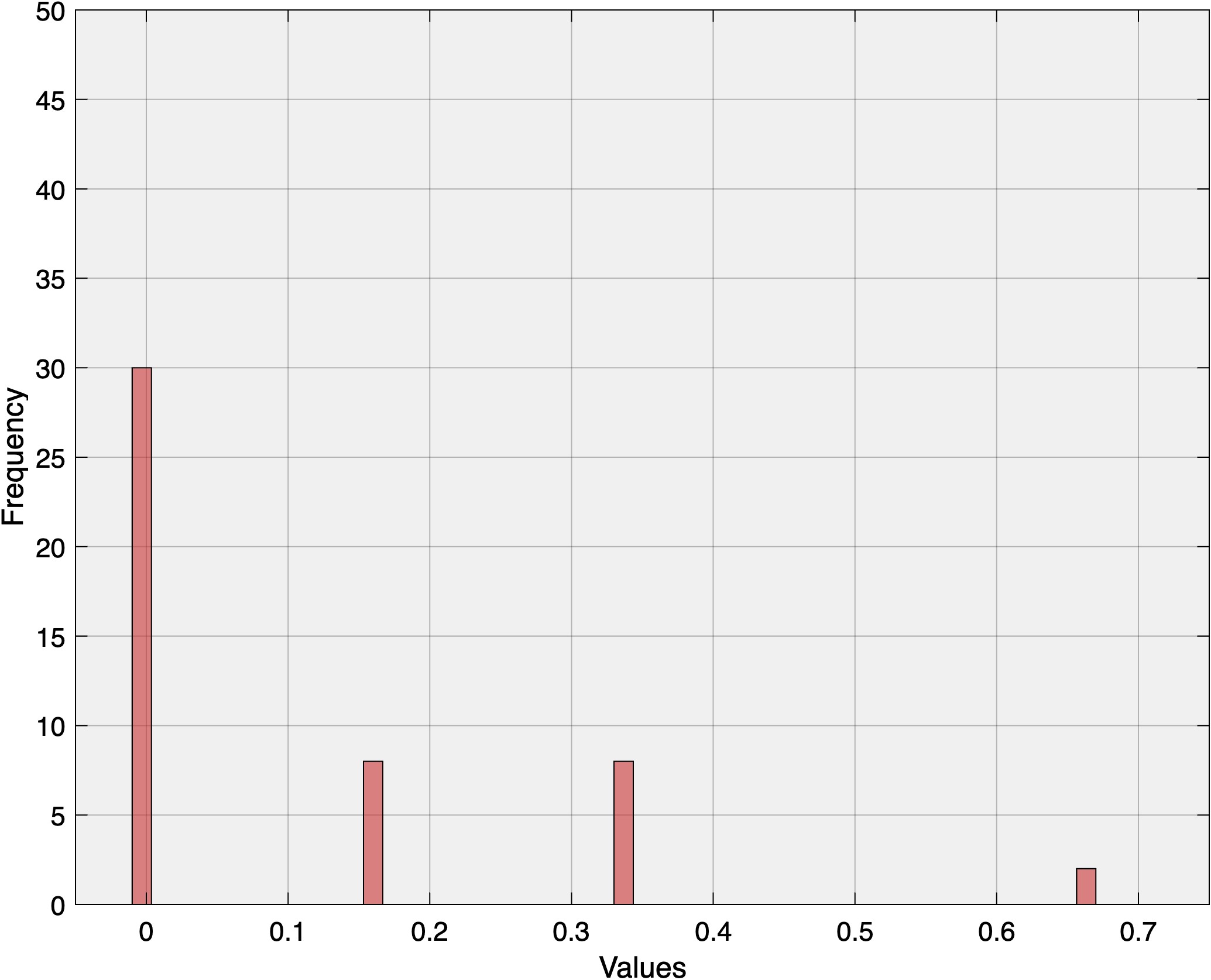}
    \caption{Frequencies of the non-zero entries of the Choi matrix where $\mathcal{S}$ consists of $N = 100$ (left) and $N = 100000$ (right) elements.}
    \label{Fibunacci_SamplingSet}
\end{figure}

Figure \ref{Fibunacci_SamplingSet} illustrates the distribution of Choi matrix entries for two sampling sizes: $|\mathcal{S}| = 100$ (left panel) and $|\mathcal{S}| = 100\,000$ (right panel). For $|\mathcal{S}| = 100$, the Choi matrix contains 45 entries exceeding $10^{-9}$, with magnitudes of approximately $10^{-3}$.
As the sampling size increases, these spurious entries diminish systematically. At $|\mathcal{S}| = 100\,000$, only 30 entries remain above the $10^{-9}$ threshold, now with magnitudes reduced to approximately $10^{-8}$
This trend demonstrates the convergence toward the exact sparse structure of the theoretical Choi matrix as the sampling becomes more comprehensive.

\subsection{Universal Symmetric Cloning}
In this section, we provide detailed numerical results for various universal symmetric $M \to N$ cloning processes. The optimal fidelity for these scenarios is analytically given by \cite{Buzek1996}:
\begin{equation}
    F^{opt}_{M \to N} = \frac{N M + M + N}{N(M + 2)}.
\end{equation}
The sampling set $\mathcal{S}$ is constructed differently depending on the number of input copies: for $M=1$, we employ the SIC-POVM states defined in Eq.~(\ref{eqn:sic-povm}); for $M=2$, we use the states given in Eq.~(\ref{eqn:pauli_set}); and for $M=3$, we generate a Fibonacci lattice sampling comprising $250\,000$ states on the appropriate Bloch sphere.

Analysis of the computed Choi matrices reveals that several exhibit numerous eigenvalues close to zero (on the order of $10^{-10}$). This leads to an excessive number of Kraus operators being extracted during the decomposition. To address this numerical artifact, we impose a tolerance threshold of $10^{-7}$ for the eigenvalue cut-off in the Kraus decomposition. With this threshold, the completeness relation $\sum_k K_k^\dagger K_k = \mathbb{1}$ remains satisfied to within approximately $10^{-8}$. Table \ref{tab:MtoNcloner} shows the results obtained.

\begin{table}[h]
    \centering
    \caption{Numerical results for universal symmetric cloning. The \textit{Gap} represents the difference between the primal and dual objective values. \textit{\#Op.} indicates the number of Kraus operators extracted from the optimal Choi matrix (after thresholding eigenvalues). The last column lists the average computation time (in seconds) for the primal solution on standard hardware.}
    \label{tab:MtoNcloner}
    \begin{tabular}{llccccr}
        \toprule
        $M \to N$ & Theoretical & Primal & Dual & Gap & \#Op. & Time \\
        \midrule
        $1 \to 2$ & $\frac{5}{6}   \approx 0.8333333333$   & 0.8333333328 & 0.8333333354 & 2.54e-09 &  2 & 0.18 \\[1ex]
        $1 \to 3$ & $\frac{7}{9}   \approx 0.7777777778$   & 0.7777777777 & 0.7777777778 & 1.42e-10 &  3 & 0.21 \\[1ex]
        $1 \to 4$ & $\frac{3}{4}   = 0.75$                 & 0.7499999925 & 0.7500000016 & 9.07e-09 &  4 & 0.24 \\[1ex]
        $1 \to 5$ & $\frac{11}{15} \approx 0.7333333333$   & 0.7333333285 & 0.7333333285 & 6.40e-09 &  5 & 0.63  \\[1ex]
        $1 \to 6$ & $\frac{13}{18} \approx 0.7222222222$   & 0.7222222196 & 0.7222222238 & 4.15e-09 &  6 & 4.62  \\[1ex]
        $1 \to 7$ & $\frac{15}{21} \approx 0.7142857143$   & 0.7142857087 & 0.7142857154 & 6.65e-09 &  7 & 40.20 \\[1ex] \hline \\    
        
        $2 \to 3$ & $\frac{11}{12} \approx 0.9166666667$   & 0.9166666605 & 0.9166666698 & 9.29e-09 & 10 & 0.32 \\[1ex] 
        $2 \to 4$ & $\frac{14}{16} = 0.875$                & 0.8749999998 & 0.8750000078 & 7.96e-09 & 19 & 0.62 \\[1ex]
        $2 \to 5$ & $\frac{17}{20} = 0.850$                & 0.8499999987 & 0.8500000002 & 1.46e-09 & 36 & 3.02 \\[1ex]
        $2 \to 6$ & $\frac{20}{24} \approx 0.8333333333$   & 0.8333333332 & 0.8333333426 & 9.42e-09 & 69 & 127.83 \\[1ex] \hline \\  
        
        $3 \to 4$ & $\frac{19}{20} = 0.95$                 & 0.9499999935 & 0.9500000176 & 2.41e-08 & 66 & 9.37 \\[1ex]
        $3 \to 5$ & $\frac{23}{25} = 0.92$                 & 0.9199999994 & 0.9200000001 & 6.51e-10 & 131& 123.75\\[1ex]
        \bottomrule
    \end{tabular}
\end{table}

\subsection{Universal Covariant Symmetric Cloning}
For phase-covariant symmetric $1 \to N$ cloning, the optimal fidelity is analytically given by \cite{Bruss2000}:
\begin{equation}
    F^{opt} = 
    \begin{cases}
        \dfrac{1}{2} + \dfrac{\sqrt{N(N+2)}}{4N}, & N \text{ is even}, \\[6pt]
        \dfrac{1}{2} + \dfrac{(N+1)}{4N}, & N \text{ is odd}.
    \end{cases}
\end{equation}
Table \ref{tab:covariant1toNcloning}  lists the calculated results. In all cases the the SIC-POVM states defined in Eq.~(\ref{eqn:sic-povm}) are used to construct $\Omega$. Here  a tolerance threshold of $10^{-8}$ is used for the eigenvalue cut-off in the Kraus decomposition.

\begin{table}[h]
    \centering
    \caption{Numerical results for phase-covariant symmetric $1 \to N$ cloning. The \textit{Gap} is the difference between the primal and dual solutions. \textit{\#Op.} denotes the number of extracted Kraus operators. The last column indicates the average computation time for the primal problem.}
    \label{tab:covariant1toNcloning}
    \begin{tabular}{llccccr}
        \toprule
        $1 \to N$ & Theoretical & Primal & Dual & Gap &  \#Op. & Sec. \\
        \midrule
        $1 \to 2$ & $ 0.8535533906$ & 0.8535533901 & 0.8535533925 & 2.437e-09 &  2 & 0.19 \\[1ex]
        $1 \to 3$ & $ 0.8333333333$ & 0.8333333292 & 0.8333333337 & 4.486e-09 &  1 & 0.21 \\[1ex]
        $1 \to 4$ & $ 0.8061862178$ & 0.8061862177 & 0.8061862217 & 3.986e-09 &  2 & 0.25 \\[1ex]
        $1 \to 5$ & $ 0.8000000000$ & 0.7999999866 & 0.8000000002 & 1.353e-08 &  3 & 0.62 \\[1ex]
        $1 \to 6$ & $ 0.7886751346$ & 0.7886751344 & 0.7886751347 & 3.165e-10 &  2 & 5.00 \\[1ex]
        $1 \to 7$ & $ 0.7857142857$ & 0.7857142844 & 0.7857142864 & 2.052e-09 &  3 & 44.52 \\[1ex]
        \bottomrule
    \end{tabular}
\end{table}

\subsection{Kraus Operators for Entanglement Cloning} \label{app:entanglement_kraus}
\begin{figure}[h]
    \centering
    \includegraphics[height=6cm]{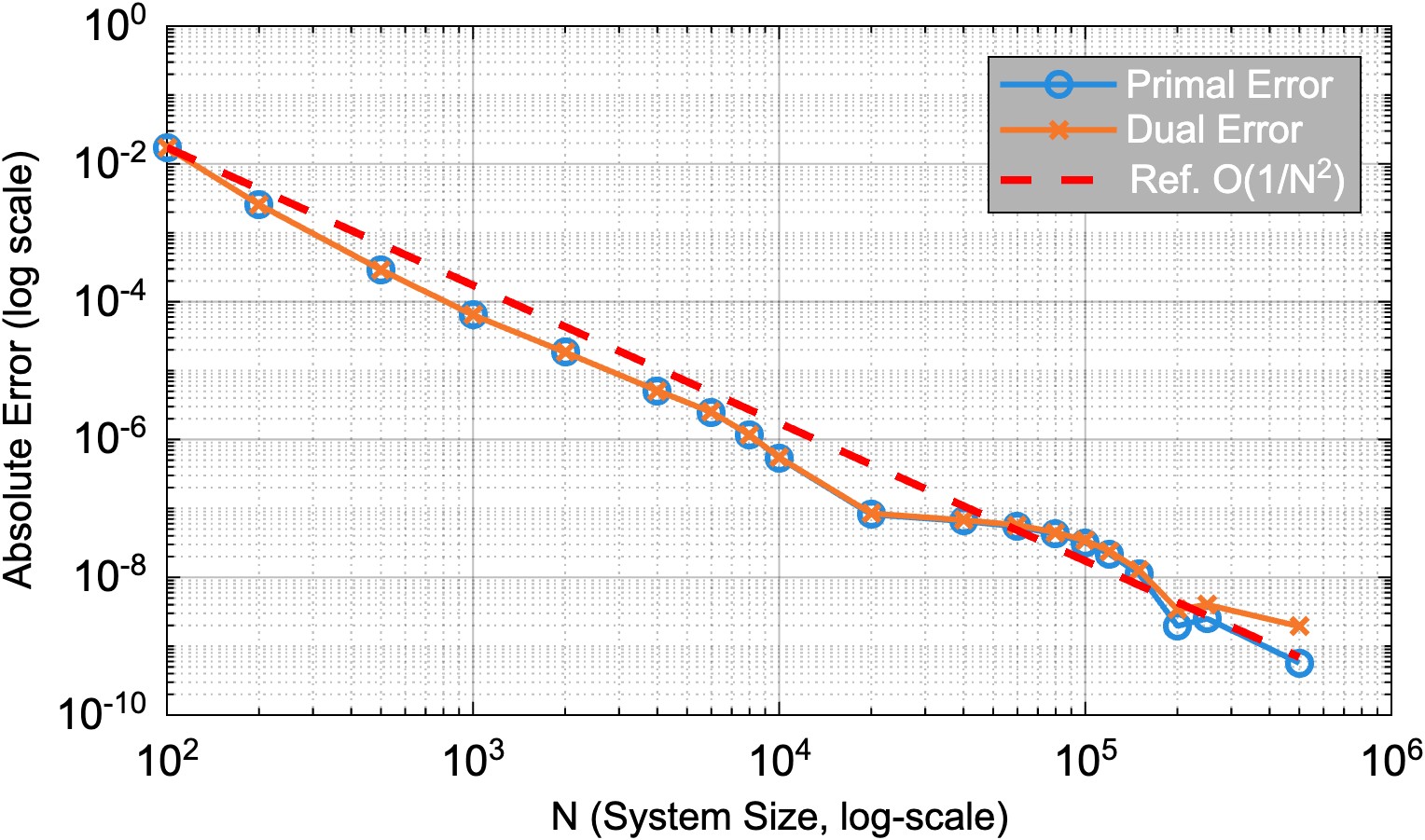}
    \caption{Convergence of the numerical error ($|F_{opt} - primal|$ and $|F_{opt} - dual|$) for the entanglement cloner as a function of the sampling set size $N$.}
    \label{Convergence_Entanglement_Analysis}
\end{figure}

For the general case of entanglement cloning, we sampled $|\cal S|$ to construct $\Omega$ with varying set sizes $|\mathcal{S}|$, as no minimal informationally complete sets are known for this specific task (general maximal entangled two qubit states). Table \ref{tab:entaglementcloning} summarizes the numerical results for the $2 \to 4$ entanglement cloner. For sample sizes exceeding $10,000$, the SDP consistently identifies four dominant Kraus operators. The convergence behaviour of the fidelity error relative to the sample size is illustrated in Fig. \ref{Convergence_Entanglement_Analysis}.

\begin{table}[h]
    \centering
    \caption{Numerical results for the $2 \to 4$ entanglement cloner across different sample sizes. The theoretical fidelity is $F_{\text{opt}} = (5 + \sqrt{13})/12 \approx 0.7171292$. Gaps represent the deviation of primal and dual solutions from this analytical bound. }
    \label{tab:entaglementcloning}
    \begin{tabular}{lcccc}
        \toprule
        No. states & Primal & Primal Gap & Dual & Dual Gap  \\
        \midrule
        100     & 0.7344266708 & -1.730e-02 & 0.7344267009 & 1.730e-02  \\[1ex]
        1000    & 0.7171933328 & -6.406e-05 & 0.7171933478 & 6.407e-05  \\[1ex]
        10000   & 0.7171298139 & -5.409e-07 & 0.7171298218 & 5.489e-07  \\[1ex]
        100000  & 0.7171293048 & -3.180e-08 & 0.7171293071 & 3.412e-08  \\[1ex]
        200000  & 0.7171292749 & -1.964e-09 & 0.7171292763 & 3.394e-09  \\[1ex]
        500000  & 0.7171292735 & -5.719e-10 & 0.7171292749 & 1.943e-09  \\[1ex]
        \bottomrule
    \end{tabular}
\end{table}

When the task is restricted to maximally entangled states with real amplitudes, a smaller sampling set of only eight states is sufficient to achieve convergence to the optimal fidelity $F_{\text{opt}} = \frac{1}{2} + \frac{1}{2\sqrt{2}}$ 
(see Sec.~\ref{sec:entanglementConling}). In this case, the channel is fully described by four Kraus operators. Defining the constants:
\begin{equation}
    a = \frac{1}{4 \sqrt{2}} \approx 0.176776, \quad b = \frac{1}{4} + \frac{1}{4 \sqrt{2}} \approx 0.426776, \quad c = \frac{1}{4} - \frac{1}{4 \sqrt{2}} \approx 0.0732233
\end{equation}
the resulting Kraus operators $K_i \in \mathbb{C}^{16 \times 4}$ can be expressed as:
\begin{equation} \scriptsize
    \begin{pmatrix}
        0  &  0  &  0  &  b  \cr
        0  &  0  & -a  &  0  \cr
        0  & -a  &  0  &  0  \cr
        b  &  0  &  0  &  0  \cr
        0  &  0  & -a  &  0  \cr 
        0  &  0  &  0  &  c  \cr
        -c  &  0  &  0  &  0  \cr
        0  &  a  &  0  &  0  \cr
        0  & -a  &  0  &  0  \cr
        -c  &  0  &  0  &  0  \cr
        0  &  0  &  0  &  c  \cr
        0  &  0  &  a  &  0  \cr
        b  &  0  &  0  &  0  \cr
        0  &  a  &  0  &  0  \cr
        0  &  0  &  a  &  0  \cr
        0  &  0  &  0  &  b
    \end{pmatrix}, \quad
    \begin{pmatrix}
        0 &  0  &  c  &  0 \cr
        0 &  0  &  0  & -a \cr
        a &  0  &  0  &  0 \cr
        0 & -c  &  0  &  0 \cr
        0 &  0  &  0  & -a \cr
        0 &  0  &  b  &  0 \cr
        0 &  b  &  0  &  0 \cr
        -a &  0  &  0  &  0 \cr
        a &  0  &  0  &  0 \cr
        0 &  b  &  0  &  0 \cr
        0 &  0  &  b  &  0 \cr
        0 &  0  &  0  &  a \cr
        0 & -c  &  0  &  0 \cr
        -a &  0  &  0  &  0 \cr
        0 &  0  &  0  &  a \cr
        0 &  0  &  c  &  0
    \end{pmatrix},\quad
    \begin{pmatrix}
        0	& -c  &	 0	&  0  \cr
        -a	&  0  &	 0	&  0  \cr
        0	&  0  &	 0	&  a  \cr
        0	&  0  &	 c	&  0  \cr
        -a	&  0  &	 0	&  0  \cr
        0	& -b  &	 0	&  0  \cr
        0	&  0  &	-b	&  0  \cr
        0	&  0  &	 0	& -a  \cr
        0	&  0  &	 0	&  a  \cr
        0	&  0  &	-b	&  0  \cr
        0	& -b  &	 0	&  0  \cr
        a	&  0  &	 0	&  0  \cr
        0   &  0  &	 c	&  0  \cr
        0	&  0  &	 0	& -a  \cr
        a	&  0  &	 0	&  0  \cr
        0	& -c  &	 0	&  0  \cr
    \end{pmatrix}, \quad
    \begin{pmatrix}
        b &  0 &  0 &  0  \cr
        0 &  a &  0 &  0  \cr
        0 &  0 &  a &  0  \cr
        0 &  0 &  0 &  b  \cr
        0 &  a &  0 &  0  \cr
        c &  0 &  0 &  0  \cr
        0 &  0 &  0 & -c  \cr
        0 &  0 & -a &  0  \cr
        0 &  0 &  a &  0  \cr
        0 &  0 &  0 & -c  \cr
        c &  0 &  0 &  0  \cr
        0 & -a &  0 &  0  \cr
        0 &  0 &  0 &  b  \cr
        0 &  0 & -a &  0  \cr
        0 & -a &  0 &  0  \cr
        b &  0 &  0 &  0  \cr
    \end{pmatrix}
\end{equation}
The Kraus operators satisfy the completeness relation.

\end{appendices}

\end{document}